\documentclass[12pt]{article}
\usepackage{comment}
\usepackage[square,comma,numbers,sort&compress]{natbib}
\usepackage[left=20mm,right=20mm,top=20mm,bottom=20mm]{geometry}

\usepackage{mediabb}
\usepackage{graphicx}

\usepackage{feynmp}
\usepackage{epsf}
\usepackage{color}
\usepackage{amsmath}
\usepackage{amssymb}
\usepackage{latexsym}
\usepackage{slashed}
\usepackage{float}
\usepackage{cases}
\usepackage{multirow}
\usepackage{bm}

\newcommand{\vev}[1]{\left\langle#1\right\rangle}

\newcommand{\A}{\mathcal{A}}

\newcommand{\D}{\mathcal{D}}
\newcommand{\E}{\mathcal{E}}

\newcommand{\U}{\mathcal{U}}

\newcommand{\Y}{\mathcal{Y}}

\newcommand{\al}[1]{\begin{align}#1\end{align}}

\newcommand{\bp}{\begin{pmatrix}}
\newcommand{\ep}{\end{pmatrix}}
\newcommand{\bb}{\begin{bmatrix}}
\newcommand{\eb}{\end{bmatrix}}

\newcommand{\paren}[1]{\left(#1\right)}
\newcommand{\sqbr}[1]{\left[#1\right]}
\newcommand{\ab}[1]{\left|#1\right|}

\newcommand{\br}[1]{\left\{#1\right\}}

\newcommand{\te}{\text}

\newcommand{\beq}{\begin{equation}}
\newcommand{\eeq}{\end{equation}}
\newcommand{\bea}{\begin{eqnarray}}
\newcommand{\eea}{\end{eqnarray}}

\newcommand{\pal}{\partial}

\newcommand{\fulltoday}{\number\day\space \ifcase\month\or
    January\or February\or March\or April\or May\or June\or
    July\or August\or September\or October\or November\or December\fi
    \space\number\year}

\begin{document}

%\maketitle
\begin{titlepage}
\renewcommand\thefootnote{\alph{footnote}}
		\mbox{}\hfill KOBE-TH-12-04\\
		\mbox{}\hfill HRI-P-12-09-001\\
		\mbox{}\hfill RECAPP-HRI-2012-010\\
\vspace{4mm}
\begin{center}
{\fontsize{22pt}{0pt}\selectfont \bf Quark mass hierarchy and mixing via {geometry}\\ of extra dimension with {point interactions}} \\
\vspace{8mm}
	\large
	Yukihiro Fujimoto,\footnote{
		E-mail: \tt 093s121s@stu.kobe-u.ac.jp
		}
	{}
%	\Large
	Tomoaki Nagasawa,\footnote{
		E-mail: \tt nagasawa@gt.tomakomai-ct.ac.jp
		}\smallskip\\
	{}
%	\Large
	Kenji Nishiwaki,\footnote{
		E-mail: \tt nishiwaki@hri.res.in
		}
%	\Large
%	Satoshi Ohya,\footnote{
%		E-mail: \tt ohya@mri.ernet.in
%	}
	{} and
	Makoto Sakamoto\footnote{
		E-mail: \tt dragon@kobe-u.ac.jp} \\
\vspace{8mm}
%	\medskip\\
	{\fontsize{14pt}{0pt}\selectfont
		${}^{\mathrm{a,d}}$\it Department of Physics, Kobe University, Kobe 657-8501, Japan \smallskip\\
		${}^{\mathrm{b}}$\it Tomakomai National College of Technology, 443 Nishikioka, Tomakomai 059-1275, Japan \smallskip\\
		${}^{\mathrm{c}}$\it Regional Centre for Accelerator-based Particle Physics, \\
		\it Harish-Chandra Research Institute, Allahabad 211 019, India
		\smallskip\\
%		${}^{\mathrm{d}}$\it Harish-Chandra Research Institute, Chhatnag Road, Jhusi,\\ Allahabad 211 019, India
%		\smallskip\\
	}
\vspace{4mm}
\end{center}
\begin{abstract}
{\fontsize{12pt}{16pt}\selectfont
We propose a new model which can simultaneously and naturally explain the origins of fermion
generation, quark mass hierarchy, and the Cabibbo--Kobayashi--Maskawa matrix
from the geometry of an extra dimension.
We take the extra dimension to be an interval with point interactions, which
are additional boundary points in the bulk space of the interval. Because of the Dirichlet boundary condition for fermions at the positions of point interactions,
profiles of chiral fermion zero modes are split and localized,
and then we can realize three generations from each five-dimensional
Dirac fermion.
Our model allows fermion flavor mixing but the form of the non-diagonal
elements of fermion mass matrices is found to be severely restricted
due to the geometry of the extra dimension.
The Robin boundary condition for a scalar leads to an extra
coordinate-dependent vacuum expectation value, which can naturally
explain the fermion mass hierarchy.
}
\end{abstract}
%\vfill
%		\mbox{}\hfill KOBE-TH-??\\
%		\mbox{}\hfill HRI-P-12-??-???\\
%		\mbox{}\hfill RECAPP-HRI-2012-???
\end{titlepage}
\renewcommand\thefootnote{\arabic{footnote}}
\setcounter{footnote}{0}
\section{Introduction}
%%%%%%%%%%%%%%%%%%%%%%%%
%%%%%%%%%%%%%%%%%%%%%%%%%%%%%%%%%%%%%%%%%%%%%

Recently{,} the ATLAS and CMS experimental groups of the CERN Large Hadron Collider (LHC) have announced the excess at $125\,\text{GeV}$, which is consistent with the Standard Model (SM) Higgs boson, with a local significance of $5 \sigma$ after combining $7\,\text{TeV}$ and $8\,\text{TeV}$ data~\cite{:2012gk, :2012gu}.
This amazing {happening means that the} mysteries behind the last missing piece of the SM are ready to be unveiled.
But the SM still has many points which are unclear{,} in spite of lots of effort {from} physicists.

One is called {the} ``quark mass hierarchy problem".
In the SM, we are forced to comply with {a} hierarchy of almost five orders of magnitude in Yukawa couplings of quarks for describing the suitable quark masses.
Closely related to this issue, the SM cannot answer the mechanism behind the {Cabibbo--Kobayashi--Maskawa} (CKM) matrix, which describes the strengths of generation-changing interactions in the SM.
In addition to {these} two issues, we cannot explain why we introduce three copies of quarks whose quantum numbers are the same except {for} their masses and the degrees of mixings in the above interactions.
Many attempts have been {made to explain} the issues within {the four-dimensional} (4D) Quantum Field Theory (QFT) framework with, {including, for example,} launching new continuous and/or discrete {symmetries}, introducing new {matter} and interactions, {and} discussing renormalization group (RG) effects from a theory {at} a {(very)} high energy scale compared to the {electroweak (EW)} scale.

When we focus on the case in five {dimensions} (5D), where {there} is one additional {spatial} direction, we can find a new useful tool for tackling the above and {other problems:} {\it geometry}.
Two of the most renowned studies which show the power of geometry are~\cite{ArkaniHamed:1998rs, Randall:1999ee}, where the authors proposed innovative ways for solving {the} hierarchy problem.
Extra space can have a huge variety of structure, which are detected as differences {from the} 4D effective theory point of view.
In {a} 5D QFT framework, we also find new mechanisms which we cannot find in 4D, {for example},
generating spontaneous gauge symmetry breaking {with a} global Wilson loop operator~\cite{Hosotani:1983xw,Hosotani:1988bm,Hatanaka:1998yp}, 
{and} symmetry breaking by orbifold boundary conditions (BCs)~\cite{Kawamura:1999nj,Kawamura:2000ev,Hebecker:2001jb}.
{However, it remains difficult to determine} the origins of the quark mass
hierarchy, {quark mixing}{,} and the number of fermion {generations based only} on them.
We {note} that the existence of (compact) extra dimension(s) is suggested by {superstring theory}.
{Lots of work has been done towards} settling the three problems in the quark (and lepton sectors) independently and/or simultaneously in the many contexts of
{large extra dimension~\cite{Dvali:1999cn,Mohapatra:1999zd,Yoshioka:1999ds,ArkaniHamed:1998vp}},
warped extra dimension~\cite{Grossman:1999ra,Gherghetta:2000qt,Huber:2000ie},
%kink profile~\cite{Grossman:2002pb,Grossman:2004rm,Melfo:2006hh,Dzhunushaliev:2008zz,Davies:2007xr},
vortex profile~\cite{Libanov:2000uf,Frere:2000dc,Frere:2001ug} based on~\cite{Nielsen:1973cs},
and others~\cite{Neronov:2001qv,Gogberashvili:2007gg,Kaplan:2011vz}.
%introducing bulk mass~\cite{Hebecker:2002re,Kakizaki:2001ue,Biggio:2003kp,Agashe:2008fe,Watanabe:2009br} 

In this paper, we focus on one of the interesting {mechanisms} {resulting from an} extra dimension, {i.e.}, localization of fields.
We can generate the hierarchy in the Yukawa coupling naturally when the SM fermions are localized at different points in one (more) extra dimension(s)~\cite{ArkaniHamed:1999dc}, whose situation is realized by a 5D scalar field coupling to 5D fermions with a kink background~\cite{Rubakov:1983bb, Akama:1982jy}.
A variation of this possibility is to {localize} the Higgs scalar vacuum expectation value (VEV) in one extra dimension~\cite{Dvali:2000ha, Kaplan:2001ga}.
Here{,} we propose a simple way of realizing three chiral generations and their localization, where we introduce point interactions {(or many branes)} on an interval~{\cite{Hatanaka:1999ac,Nagasawa:2002un,Nagasawa:2003tw,Nagasawa:2005kv}}.
This system is decomposed into {multiple intervals and,}
due to the Dirichlet BC at the positions of {the} point interactions, {fermion zero modes are split and degenerated.}
Each profile is localized around a
corresponding {point interaction} as an effect of nonzero fermion bulk mass.

When we construct a model with the above mechanism, {it is reasonable to assume {that} all the 5D fields live in the bulk with no tree-level localized term at the positions of point interactions}.
This setup is very similar to the {minimal Universal Extra Dimension (mUED)} model on $S^1/Z_2$~\cite{Appelquist:2000nn}.
{The} mUED is one of the most investigated {models} in {the context of an} extra dimension and has many {exciting} points, {e.g.}, {the existence of a} dark matter candidate is ensured by the accidental symmetry under changing the two end points of $S^1/Z_2$~\cite{Servant:2002aq}.\footnote{
{Recently, a non-minimal version of {the} UED model with brane-localized terms has been proposed in Ref.~\cite{Flacke:2008ne}, and the collider physical studies on the model have also been done in Refs.~\cite{Flacke:2008ne,Datta:2012xy,Datta:2012tv,Flacke:2012ke}.}
}
The latest concrete analysis for relic abundance of the candidate is found in Ref.~\cite{Belanger:2010yx}.
The parameters of the mUED (and other {six-dimensional} UED models) are restricted by {analyses} based on {the} recent LHC experimental results in Refs~\cite{Nishiwaki:2011vi,Nishiwaki:2011gk,Belanger:2012sy,Kakuda:2012px,Belanger:2012mc}.
In the mUED model, the Yukawa structure is the same {as} that of the SM, and therefore we still need {some} fine tuning in the coefficients.
{On the other hand, in our model}, the 5D Yukawa couplings cannot possess {a} generation index since {one generation} of the SM fields {is} introduced.
In other words, another maneuver should be offered to overwhelm the difficulty.

To generate the Yukawa {coupling} hierarchy via geometry, an extra coordinate-dependent and localized profile of the {scalar} VEV is preferred.
An idea to realize this situation is to impose nontrivial BCs for the scalar field which is incompatible with its non-vanishing constant vacuum configuration.
{This mechanism has been applied to breaking translational invariance~\cite{Sakamoto:1999yk},
breaking supersymmetry~\cite{Sakamoto:1999ym, Sakamoto:1999iv,Ohnishi:2000hs,Hatanaka:2000zq}{, and has been} extended to higher extra dimensions~\cite{Matsumoto:2001fp,Sakamoto:2001gn}.}
For the scalar singlet case, the profile is described with {Jacobi's elliptic function} and we can find a parameter region where the elliptic function approaches the exponential function.
The exponential form is ideal for generating {a} large hierarchy within a natural choice of parameters{,} and almost all the input parameters take coefficients with $\mathcal{O}(10)$ magnitude when we scale them based on each corresponding suitable mass value.

This paper is organized as follows:
In Section~\ref{section:basic} we give a brief review {of} a way of constructing system with many point interactions{,} {and subsequently discuss} suitable choices of BCs for {a} 5D fermion and vector fields.
In Section~\ref{assumedVEVanalysis_section} we search for the possibility {of} achieving quark mass hierarchy and mixing simultaneously in {a multiple} {point interaction} system with {an} exponential Higgs VEV profile.
In Section~\ref{concretemodel_section} we construct a concrete model realizing the exponential VEV with high precision without violating gauge coupling universality{,} and check the validity of the model through discussing {the naturalness of {the} magnitudes of the} coefficients.
Section~\ref{Summary_section} is devoted to conclusions and discussions.

%%%%%%%%%%%%%%%%%%%%%%%%%%%%%%%%%%%%%%%%%%%%%
%%%%%%%%%%%%%%%%%%%%%%%%
\section{{Basic properties of zero mode functions in {a} system with {point interactions}}
\label{section:basic}}
%%%%%%%%%%%%%%%%%%%%%%%%
%%%%%%%%%%%%%%%%%%%%%%%%%%%%%%%%%%%%%%%%%%%%%

%%%%%%%%%%%%%%%%%%%%%%%%%%%%%%%%%%%%%%%%%%%%%
\subsection{Zero mode profile of 5D fermion on an interval with {a} point interaction}
%%%%%%%%%%%%%%%%%%%%%%%%%%%%%%%%%%%%%%%%%%%%%

\begin{figure}
\begin{center}
\includegraphics[width=70mm]{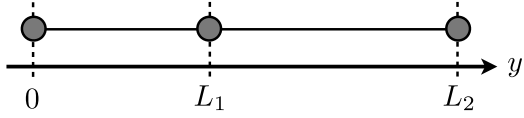}
\end{center}
\caption{{A schematic diagram of the interval system with a {point interaction} at $y=L_1$.}}
\label{twointervals_pdf}
\end{figure}

In this subsection, we consider the zero mode profile of {a} 5D fermion on an interval with {a} point interaction {which is placed} in the middle of the whole system.
{A point interaction means that the interaction can occur
only at a point as a $\delta$-function potential.
In Refs.~\cite{QMSUSY1,QMSUSY2,Cheon:2000tq,Nagasawa:2008an}, possible point interactions in
one-dimensional quantum  mechanics are shown to be
classified by boundary conditions which are characterized
by $U(2)$ parameters for a circle and $U(1) \times U(1)$
for an interval. According to this result, we will
specify each point interaction by one of the possible
boundary conditions in this paper.}
%In this paper, we use the technical term ``{point interaction}" as the meaning of additional boundary point which is located at the bulk space of the original interval.
We use a coordinate $y$ to indicate the position in the extra space and assign the locations of the three boundary points as $0\,(=L_0), L_1, L_2$, respectively.
The schematic diagram of our system in Fig.~\ref{twointervals_pdf} helps our understanding.

In this paper, we concentrate on {the simple case} where there is no tree-level brane localized term. 
{The 5D action of a fermion we consider is}
\al{
\int d^4 x {\Bigl[\int_{0}^{L_1} dy+\int_{L_1}^{L_2}dy\Bigr] }\biggl( \overline{\Psi} \paren{i \pal_M \Gamma^M + M_F} \Psi \biggr){,}
\label{fermionfreeaction}
}
where the latin indices run {from} $0$ to $3,5\ (\te{or\ }y)$ and greek ones
run {from} $0$ to $3$, respectively.\footnote{
In this paper, we choose {the} metric convention as
$
\eta_{MN} = \eta^{MN} = \text{diag}(-1,1,1,1,1).
$
The representations of {the} gamma matrices are
$
\Gamma_\mu = \gamma_\mu, \ \Gamma_y = \Gamma^y = -i \gamma^5 = \gamma^0 \gamma^1 \gamma^2 \gamma^3,
$
and we {note that} the Clifford algebra is defined as
$
\br{\Gamma_M, \Gamma_N} = -2 \eta_{MN}.
$
}
$\Psi$ and $M_F$ are a 5D Dirac fermion and its bulk mass.
We mention that 5D fermion bulk mass with the ordinary form is allowed in our system since we do not introduce orbifold $Z_2$ parity.

In what follows, we contemplate {the} profiles of fermions.
Due to {the} variational principle, the following quantities must vanish at the corresponding {boundaries:}
\al{
&\Big[ \overline{\Psi} \Gamma^{y} \delta \Psi \Big] \Big|_{y=0} =
\Big[ \overline{\Psi} \Gamma^{y} \delta \Psi \Big] \Big|_{y=L_2} = 0, \label{fermioncondition1_twointerval}\\
&\Big[ \overline{\Psi} \Gamma^{y} \delta \Psi \Big] \Big|_{y=L_1 - \varepsilon} -
\Big[ \overline{\Psi} \Gamma^{y} \delta \Psi \Big] \Big|_{y=L_1 + \varepsilon} = 0, \label{fermioncondition2_twointerval}
}
where $\varepsilon$ is an infinitesimal positive constant.
The form of {$\overline{\Psi} \Gamma^{y} \delta \Psi=0$ }can be decomposed into $\overline{\Psi}_R \delta \Psi_L = \overline{\Psi}_L \delta \Psi_R = 0$, where the 4D chirality is defined as
$\gamma^5 \Psi_{R \atop L} = \pm \Psi_{R \atop L}$.
Then, the Dirichlet BCs turn out to be consistent with Eqs.~(\ref{fermioncondition1_twointerval}) and (\ref{fermioncondition2_twointerval}), {i.e.}
%\footnote{
%It is noted that analysis in one-dimensional quantum mechanics is useful for classifying possible class of BCs. For example, the most general BC's of a wavefunction on an interval are known to be characterized by $U(1) \times U(1)$ parameters at boundary or point interaction~\cite{QMSUSY1,QMSUSY2,Cheon:2000tq,Nagasawa:2008an}.
%}
\al{
\Psi_R = 0 \quad \text{or} \quad \Psi_L = 0 \qquad \text{at} \quad y=0,\, L_1 \pm \varepsilon,\, L_2 .
}
We will, however, choose $\Psi_{R}=0$ (or $\Psi_{L}=0$) {at all the boundaries} to realize the SM chiral fermions in the zero mode sector, as we will see later.

We {note} that once the BC of a {right- (left-)handed} part of a 5D fermion $\Psi$ is determined as
\al{
\Psi_R(x,L_i) = 0 \quad \Big( \Psi_L(x,L_i) = 0 \Big),
\label{BC_of_fermion}
}
where $L_i$ shows the position of a {boundary} point, the BC of the remaining {left- (right-)handed} part is simultaneously fixed through {the} equation of motion (EOM) as
\al{
(-\pal_y + M_{{F}})\Psi_L = 0 \quad \Big( (\pal_y + M_{{F}})\Psi_R = 0 \Big) \qquad {\te{at} \quad y = 0,\, L_1 \pm \varepsilon,\, L_2}.
\label{countBC_fermion_original}
}
We can expand the 5D fermion as
\al{
\Psi(x,y) &= \Psi_R(x,y) + \Psi_L(x,y) \notag \\
&= {\sum_{n}} \Big\{ \psi_R^{(n)}(x) {f_{\Psi_{R}^{(n)}}(y)} + \psi_L^{(n)}(x) {f_{\Psi_{L}^{(n)}}(y)} \Big\}.
}
{The series} {$\{f_{\Psi_{R}^{(n)}}\}$} and {$\{f_{\Psi_{L}^{(n)}}\}$} are eigenstates of the {hermitian operators} $\D^\dagger \D$ and $\D \D^\dagger$, respectively; {$\D$ (and $\D^\dagger$)} are defined as
\beq
\left\{
\begin{array}{l}
\D := \pal_y + M_{F} \\
\D^\dagger := -\pal_y + M_{F}
\end{array}
\right.
,
\eeq
\beq
\left\{
\begin{array}{l}
\D^\dagger \D {f_{\Psi_{R}^{(n)}}(y)} = M_{\Psi_{(n)}}^2 {f_{\Psi_{R}^{(n)}}(y)} \\
\D \D^\dagger {f_{\Psi_{L}^{(n)}}(y)} = M_{\Psi_{(n)}}^2 {f_{\Psi_{L}^{(n)}}(y)}
\end{array}
\right.
,
\eeq
where $M_{\Psi_{(n)}}$ is a KK mass of the {$n$th} right/left KK mode.
This mass degeneracy is ensured by {quantum mechanical} supersymmetry (QMSUSY)~\cite{Nagasawa:2008an,Lim:2005rc,Lim:2007fy,Lim:2008hi} as
\beq
\left\{
\begin{array}{l}
\D {f_{\Psi_{R}^{(n)}}(y)} = M_{\Psi_{(n)}} {f_{\Psi_{L}^{(n)}}(y)} \\
\D^\dagger {f_{\Psi_{L}^{(n)}}(y)} = M_{\Psi_{(n)}} {f_{\Psi_{R}^{(n)}}(y)}
\end{array}
\right.
.
\label{QMSUSY_fermions}
\eeq
For zero mode ($M_{\Psi_{(0)}} = 0$), Eq.~(\ref{QMSUSY_fermions}) {takes} the simple {form:}
\beq
\left\{
\begin{array}{l}
\D {f_{\Psi_{R}^{(0)}}(y)} = 0 \\
\D^\dagger {f_{\Psi_{L}^{(0)}}(y)} = 0
\end{array}
\right.
.
\label{QMSUSY_fermionzeromode}
\eeq
{Taking account of the BCs, we find the zero mode solutions for $\Psi_{L}=0$ at $y=0, L_{1}\pm\varepsilon, L_{2}$ as (see Fig.~2)
	%%%%%%%%%%%%%%%%%%%%%%%%%%%%%%%%%%%%%%%%%%%%%%
	\beq
	f_{\Psi_{1R}^{(0)}}(y)=\left\{\begin{array}{l}
					{\mathcal{N}}_{1}\hspace{0.15em}e^{-M_{F} y}\qquad \te{for} \quad 0<y<L_{1}\\
					0	\hspace{3.5em}\qquad \te{for} \quad L_{1}<y<L_{2}
					\end{array}
				\right.
				\label{showconcreteprofile1}
	\eeq
	%%%%%%%%%%%%%%%%%%%%%%%%%%%%%%%%%%%%%%%%%%%%%%
and
	%%%%%%%%%%%%%%%%%%%%%%%%%%%%%%%%%%%%%%%%%%%%%%
	\beq
	f_{\Psi_{2R}^{(0)}}(y)=\left\{\begin{array}{l}
					0	\hspace{3.5em}\qquad \te{for} \quad 0<y<L_{1}\\
					{\mathcal{N}}_{2}\hspace{0.15em}e^{-M_{F} y}\qquad \te{for} \quad L_{1}<y<L_{2}
					\end{array}
				\right. ;
				\label{showconcreteprofile2}
	\eeq
	%%%%%%%%%%%%%%%%%%%%%%%%%%%%%%%%%%%%%%%%%%%%%%
for $\Psi_{R}=0$ at $y=0, L_{1}\pm\varepsilon, L_{2}$:
	%%%%%%%%%%%%%%%%%%%%%%%%%%%%%%%%%%%%%%%%%%%%%%
	\beq
	f_{\Psi_{1L}^{(0)}}(y)=\left\{\begin{array}{l}
					\mathcal{N}'_{1}\hspace{0.15em}e^{M_{F} y}\qquad \hspace{0.2em}\te{for} \quad 0<y<L_{1}\\
					0	\hspace{3.2em}\qquad \te{for} \quad L_{1}<y<L_{2}
					\end{array}
				\right.
				\label{showconcreteprofile3}
	\eeq
	%%%%%%%%%%%%%%%%%%%%%%%%%%%%%%%%%%%%%%%%%%%%%%
and
	%%%%%%%%%%%%%%%%%%%%%%%%%%%%%%%%%%%%%%%%%%%%%%
	\beq
	f_{\Psi_{2L}^{(0)}}(y)=\left\{\begin{array}{l}
					0	\hspace{3.2em}\qquad \te{for} \quad 0<y<L_{1}\\
					\mathcal{N}'_{2}\hspace{0.15em}e^{M_{F} y}\qquad \hspace{0.2em}\te{for} \quad L_{1}<y<L_{2}
					\end{array}
				\right. ,
				\label{showconcreteprofile4}
	\eeq
	%%%%%%%%%%%%%%%%%%%%%%%%%%%%%%%%%%%%%%%%%%%%%%
where $\mathcal{N}_{1}, \mathcal{N}_{2}, \mathcal{N}'_{1}, \mathcal{N}'_{2}$ are normalization constants, whose concrete forms are shown {later}.\\
}

\begin{figure}
\centering
\includegraphics[width=0.75\columnwidth]{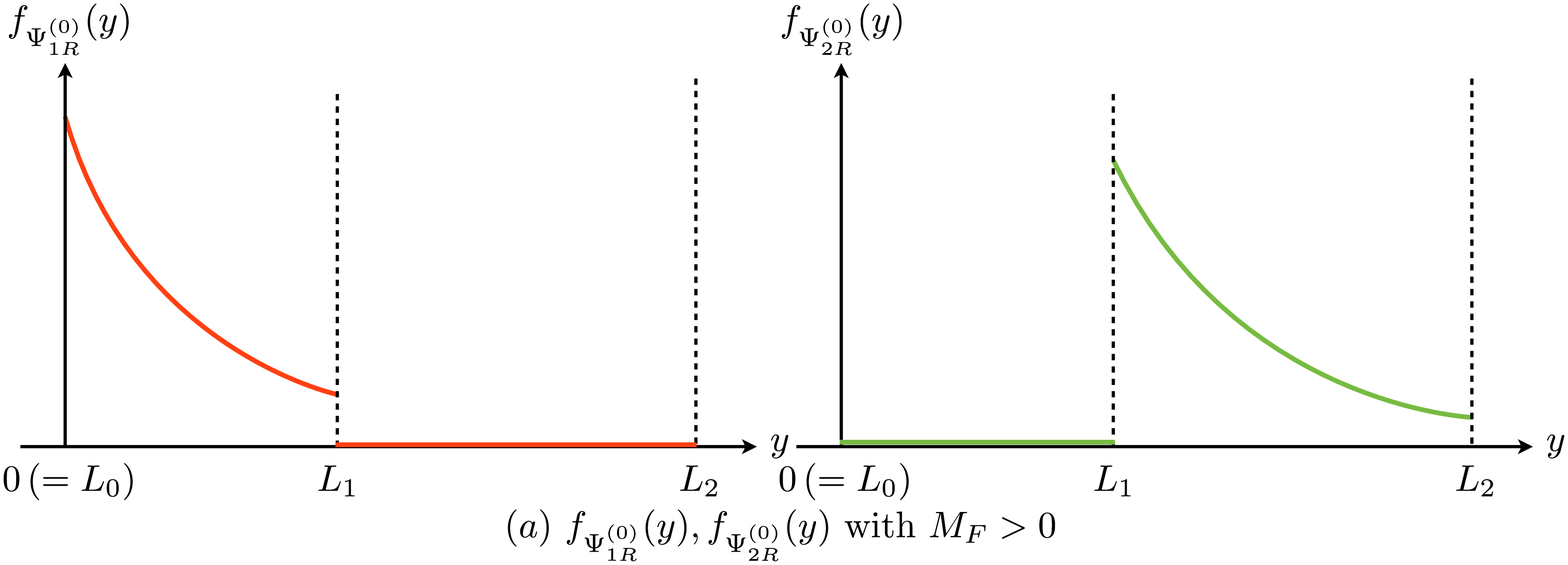} \\
\includegraphics[width=0.75\columnwidth]{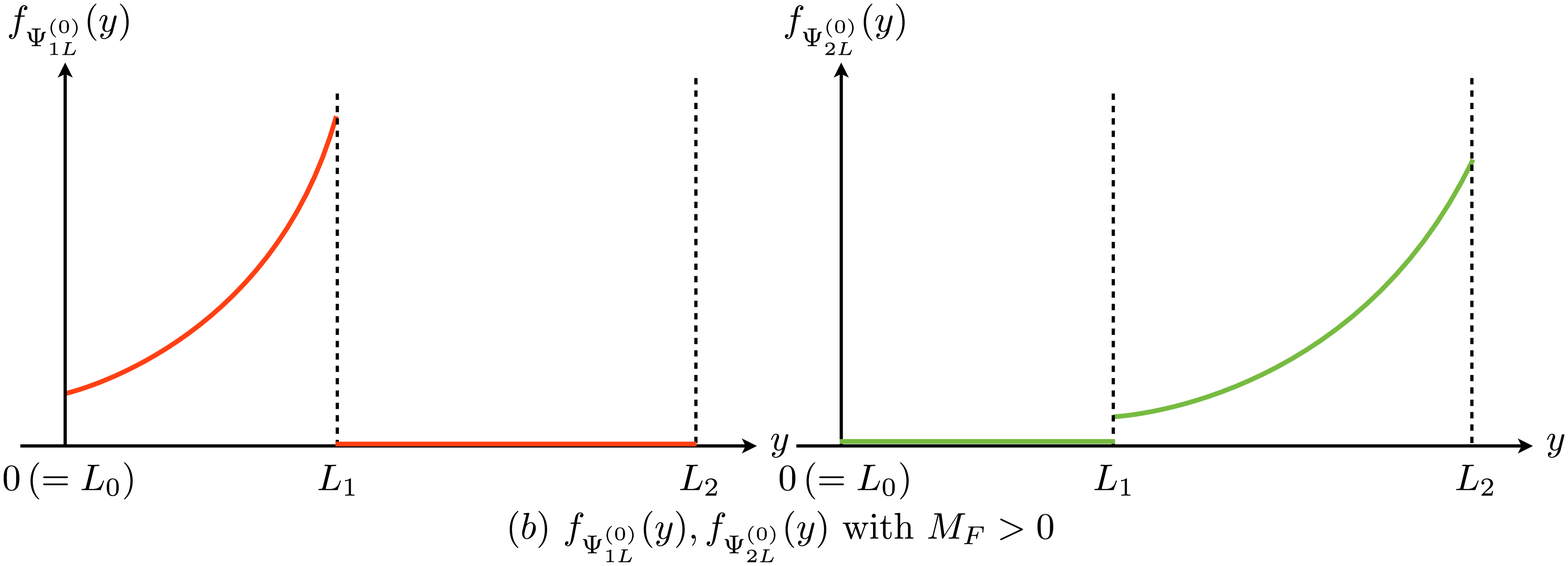}
\caption{
Profiles of the zero modes $f_{\Psi^{(0)}_{1R}}(y), f_{\Psi^{(0)}_{2R}}(y)$ {and $f_{\Psi^{(0)}_{1L}}(y), f_{\Psi^{(0)}_{2L}}(y)$ are depicted schematically in $(a)$ and $(b)$ for $\Psi_L=0$ and $\Psi_R=0$, respectively,} with $M_F>0$. Here there are {three} boundary points at {$y=0, L_1, L_2$}.
}
\label{fermionwaveprofile}
\end{figure}

{
Here{,} we would like to comment on some important points of the situation.
{The Dirichlet boundary condition (\ref{BC_of_fermion}) for {a} fermion at $y=L_{1}$} turns out to effectively split the interval into two segments.
Then, chiral zero modes are two-fold degenerate, and each profile of
the zero modes is confined in one of the segments and localized
exponentially at one of the edges,\footnote{
{It is noted that the value of $M_F$ can take a negative value{,} and in this case
the direction of wave function localization {flips} to the opposite side.}
}
as shown in Eqs. (\ref{showconcreteprofile1}) and (\ref{showconcreteprofile2}) (or (\ref{showconcreteprofile3}) and (\ref{showconcreteprofile4})).
Thus, we can realize a model with two generations of 4D chiral fermions
in the case of {an} interval with {a} {point interaction}.
Furthermore, the localization of the zero mode profiles will be found
to lead to the fermion mass hierarchy, as will {be seen} later.
We can see pictures explaining this situation in Fig.~\ref{fermionwaveprofile}. Each concrete form
is written down as follows:}

\begin{itemize}
\item In the case of $\Psi_R = 0$ at $y=0,\, L_1{\pm\varepsilon},\, L_2$,
\al{
\Psi(x,y) &= \Bigg\{ \sqrt{\frac{2M_{F}}{e^{2M_{F} \Delta L_{1}}-1}} e^{M_{F} (y-L_0)} [\theta(y-L_0) \theta(L_{1}-y)] \psi^{(0)}_{1L}(x)
\notag \\
&\phantom{=}\  + \sqrt{\frac{2M_{F}}{e^{2M_{F} \Delta L_{2}}-1}} e^{M_{F}(y-L_{1})} [\theta(y-L_{1}) \theta(L_{2}-y)] \psi^{(0)}_{2L}(x) \Bigg\}
+ (\text{KK modes}),
\label{zeromodefunctionprofile1_twointerval}
}
\item In the case of $\Psi_L = 0$ at $y=0,\, L_1{\pm\varepsilon},\, L_2$,
\al{
\Psi(x,y) &= \Bigg\{ \sqrt{\frac{2M_{F}}{1 - e^{-2M_{F} \Delta L_{1}}}} e^{-M_{F} (y-L_0)} [\theta(y-L_0) \theta(L_{1}-y)] \psi^{(0)}_{1R}(x)
\notag \\
&\phantom{=}\  + \sqrt{\frac{2M_{F}}{1 - e^{-2M_{F} \Delta L_{2}}}} e^{-M_{F}(y-L_{1})} [\theta(y-L_{1}) \theta(L_{2}-y)] \psi^{(0)}_{2R}(x) \Bigg\}
+ (\text{KK modes}),
\label{zeromodefunctionprofile2_twointerval}
}
\end{itemize}
where $\theta(y)$ is the step function and $\psi_{1}^{(0)}$ and $\psi_{2}^{(0)}$ represent two (degenerated) fermion zero modes.
$\Delta L_{i}$ shows the length of the corresponding {$i$th} segment which is defined as
\al{
\Delta L_{i} = L_{i} - L_{i-1}.
}
Here{,} every mode function is suitably normalized and we can find the factor for this purpose in front of the exponential functions.

We can also evaluate the {right- or left-handed} KK {fermion profiles}{,} but
the aim of this paper is to understand the mechanism explaining the fermion mass hierarchy.
{Therefore, we will revisit this issue in} future work.

%%%%%%%%%%%%%%%%%%%%%%%%%%%%%%%%%%%%%%%%%%%%%
\subsection{Zero mode profile of 5D gauge boson on an interval with {a} {point interaction}}
%%%%%%%%%%%%%%%%%%%%%%%%%%%%%%%%%%%%%%%%%%%%%

Following the previous section, we move to the zero mode profile of {a} 5D gauge boson on the interval with {a} {point interaction}.
Here we concentrate on {the} $U(1)$ case since our {intention} is to investigate the structure of {the} mass spectrum. The concrete form of the 5D action is
\al{
\int d^4 x{\Bigl[\int_{0}^{L_1} dy+\int_{L_1}^{L_2}dy\Bigr] }\biggl( - \frac{1}{4} F_{MN} F^{MN} \biggr) ,
}
where $F_{MN} = \pal_M A_N - \pal_N A_M$ is the 5D field strength of the 5D $U(1)$ gauge boson $A_M$ and we assume that the background geometry is the same as in the fermion case.
After taking variation, we focus on the forms of conditions at the boundary points{:}
\al{
& \Big[ (\pal_y A_\mu - \pal_\mu A_y) \delta A^{\mu} \Big] \Big|_{y=0} =
\Big[ (\pal_y A_\mu - \pal_\mu A_y) \delta A^{\mu} \Big] \Big|_{y=L_2} = 0, 
\label{gaugebosonsurfaceterm1_twointerval} \\
& \Big[ (\pal_y A_\mu - \pal_\mu A_y) \delta A^{\mu} \Big] \Big|_{y=L_1-\varepsilon} -
\Big[ (\pal_y A_\mu - \pal_\mu A_y) \delta A^{\mu} \Big] \Big|_{y=L_1+\varepsilon} = 0.
\label{gaugebosonsurfaceterm2_twointerval}
}
Under the constraints, we could choose the following BCs, where {the} Neumann (Dirichlet) type BC is assigned for $A_{\mu}$ ($A_y$), like the {mUED} model as 
\al{
(\pal_y  A_\mu) = 0 \qquad &\text{at} \quad y=0,\, L_1 \pm \varepsilon,\, L_2,
\label{gaugebosoncondition1_twointerval} \\
A_y = 0 \qquad &\text{at} \quad y=0,\, L_1 \pm \varepsilon,\, L_2,
\label{gaugebosoncondition2_twointerval}
}
in which $A_{\mu}$'s zero mode with a constant profile is found.
We {note} that the compatibility between the BCs'~(\ref{gaugebosoncondition1_twointerval}) and (\ref{gaugebosoncondition2_twointerval}) can also be shown from a viewpoint of QMSUSY~\cite{Nagasawa:2008an,Lim:2005rc,Lim:2007fy,Lim:2008hi}.
But this configuration is problematic {from} {a} phenomenological point of view.
\begin{figure}
\centering
\includegraphics[width=0.75\columnwidth]{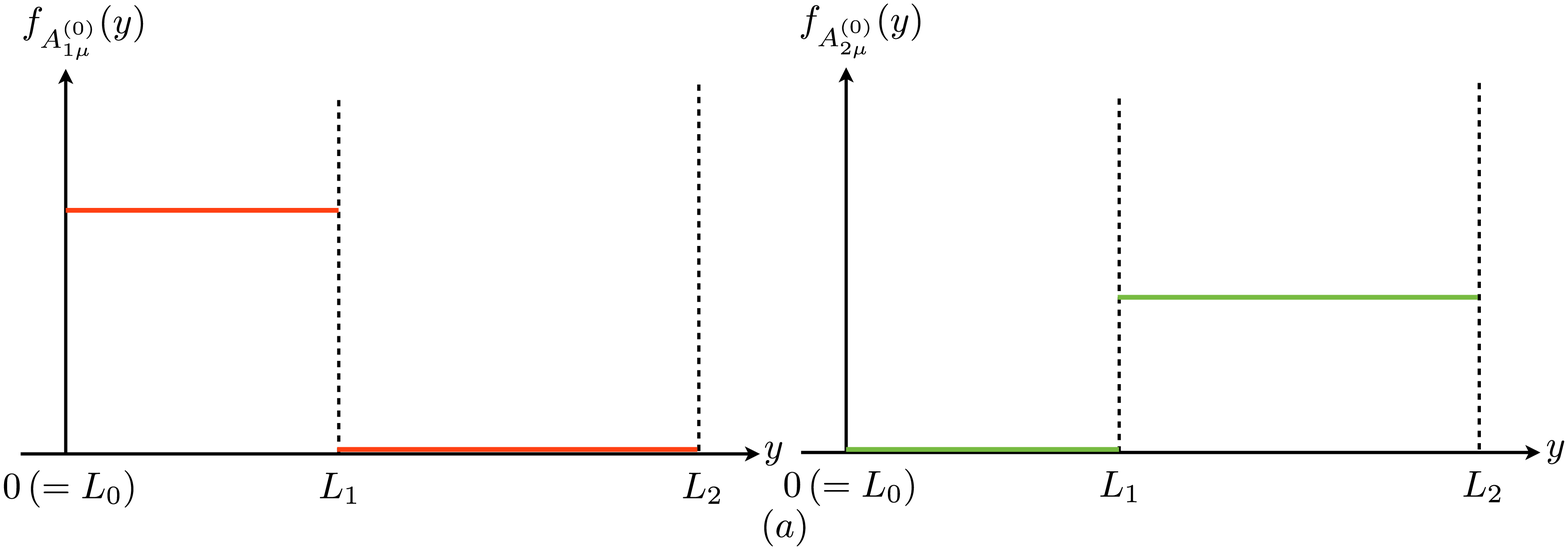} \\
\includegraphics[width=0.4\columnwidth]{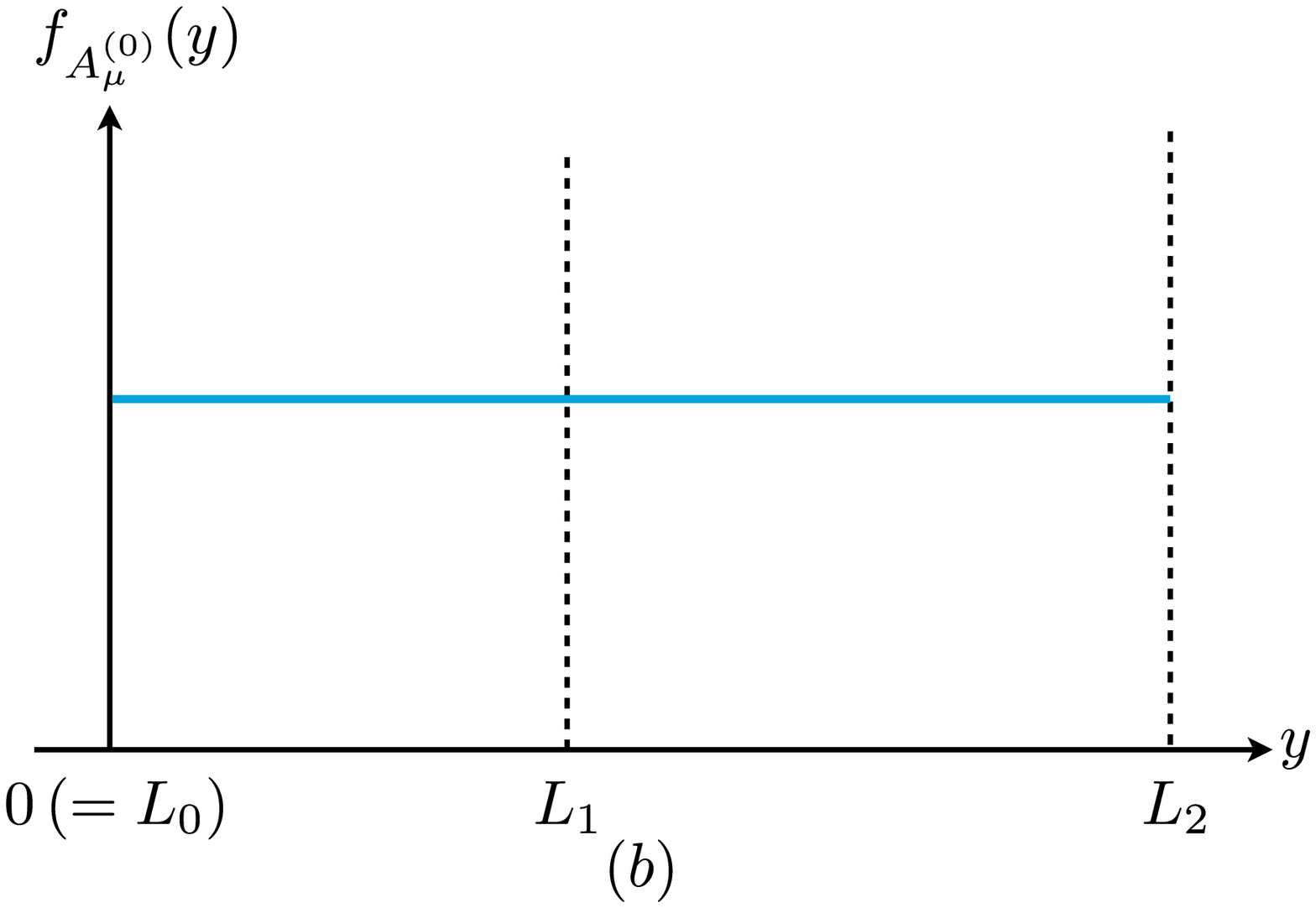}
\caption{
{{Profiles} of zero mode 4D gauge boson{: $(a)$} with BCs in Eq.~(\ref{gaugebosoncondition1_twointerval}){,
and $(b)$ with BCs} which are modified at $y=L_1$ as in Eq.~(\ref{gaugebosoncondition3_twointerval}).}
}
\label{gaugebosonwaveprofile_twointerval}
\end{figure}
In this setup, the zero mode profile of {a} 4D gauge boson is also {split at $y=L_1$} and this fact means that a zero mode within an interval {has} limited interactions only with the particles inside the {segment} {to which the gauge boson belongs} (see Fig.~\ref{gaugebosonwaveprofile_twointerval}).
Since this possibility with doubly-degenerated zero modes {(${A^{(0)}_{1\mu}}$ and ${A^{(0)}_{2\mu}}$)} is rejected in the SM, we have to change the BC at $y=L_1$.
We further {note} that the gauge universality in the SM is violated in this configuration and this fact gets to be another reason for discarding the system {with the BCs given in (\ref{gaugebosoncondition1_twointerval}) and (\ref{gaugebosoncondition2_twointerval}).}

To avoid the problems, {we would like the profile} to be continuous at $y=L_1$.
{Therefore} we put the ``continuous" conditions at this point for $A_\mu$ and $A_y$ as
\al{
A_{\mu} \Big|_{y=L_1-\varepsilon} = A_{\mu} \Big|_{y=L_1+\varepsilon} &\quad
\text{and} \quad
\pal_y A_{\mu} \Big|_{y=L_1-\varepsilon} = \pal_y A_{\mu} \Big|_{y=L_1+\varepsilon},
\label{gaugebosoncondition3_twointerval} \\
A_{y} \Big|_{y=L_1-\varepsilon} = A_{y} \Big|_{y=L_1+\varepsilon} &\quad
\text{and} \quad
\pal_y A_{y} \Big|_{y=L_1-\varepsilon} = \pal_y A_{y} \Big|_{y=L_1+\varepsilon}.
\label{gaugebosoncondition4_twointerval}
}
These conditions are consistent with the constraints {in Eq.~(\ref{gaugebosonsurfaceterm2_twointerval})}.
{In this case, we can observe only one zero mode of $A_{\mu}$, whose situation is consistent with that of the SM (see Fig.~\ref{gaugebosonwaveprofile_twointerval}).}
It is noted that we need to put the continuity in the first derivative level because the {Klein--Gordon} (or Maxwell) equation is second order.
We also consider this continuous condition for 5D scalar as in Eq.~(\ref{gaugebosoncondition4_twointerval}).

%%%%%%%%%%%%%%%%%%%%%%%%%%%%%%%%%%%%%%%%%%%%%
\subsection{Flavor mixing in {a} system with {point interactions}}
%%%%%%%%%%%%%%%%%%%%%%%%%%%%%%%%%%%%%%%%%%%%%

\begin{figure}
\centering
\includegraphics[width=0.6\columnwidth]{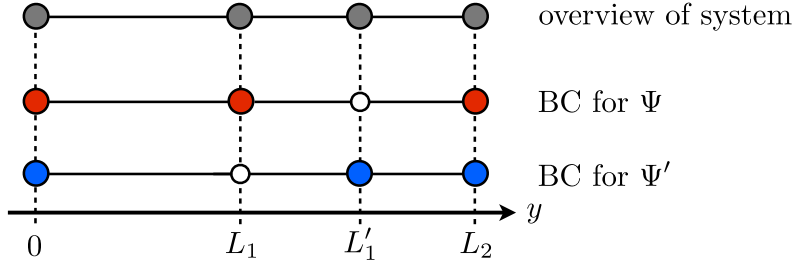}
\caption{
This is an overview of our system with two {point interactions}, where the red (blue) circular spots show the Dirichlet-{type} BC for {the} left- (right-)handed part at the corresponding boundary points, respectively, and the small white dots mean the continuous condition for fermions.}
\label{overview_threeinterval}
\end{figure}

\begin{figure}
\centering
\includegraphics[width=0.5\columnwidth]{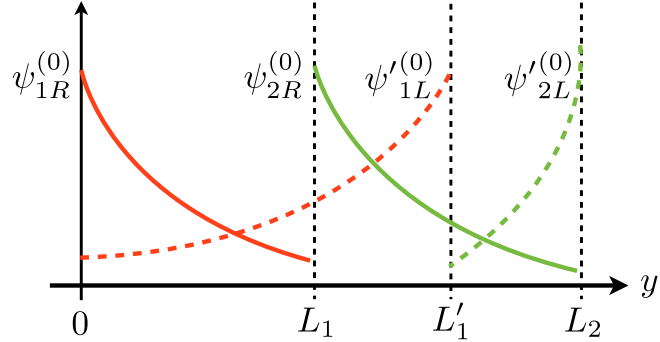}
\caption{
Zero mode profiles of 5D fermions $\Psi$ and $\Psi'$.
}
\label{fermionwaveprofile_threeinterval}
\end{figure}

We can also consider the ``continuous" condition at {point interactions} for fermions as
\al{
\Psi \Big|_{y=L_i-\varepsilon} = \Psi \Big|_{y=L_i+\varepsilon}{,}
}
{where $L_{i}$ denotes a position of point interactions.}
An interesting application is given as follows:
At this time, we add another 5D fermion $\Psi'$ with the different bulk mass $M'_F$ and go to a two-{point-interaction} system, where the {point interactions} are located at $y=L_1, L'_1$ (see Fig.~\ref{overview_threeinterval}).
The 5D action which now we think about is
\al{
\int d^4 x \int_{0}^{L_2} dy \Bigg\{ \Big[ \overline{\Psi} \paren{i \pal_M \Gamma^M + M_F} \Psi \Big] +
\Big[ \overline{\Psi'} \paren{i \pal_M \Gamma^M + M'_F} \Psi' \Big] \Bigg\},
\label{fermionfreeaction_threeinterval}
}
where we do not divide the range of integral for $y$ for clarity {of} description.
The BCs for $\Psi$ and $\Psi'$ are selected as in Fig.~\ref{overview_threeinterval}, where the red (blue) circular {spots show} the Dirichlet-{type} BC for {the} left- (right-)handed part {at the corresponding boundary points}, respectively, and the small white {dots mean the} continuous condition for {fermions}.
Under the BCs, zero modes of both $\Psi$ and $\Psi'$ become two-fold degenerated and we distinguish the two states by adding new indices of $``1"$ or $``2"$ showing ``generation" from the left to the right. 
When we choose the signs of the two bulk masses $M_F, M'_F$ as $M_F > 0, M'_F > 0$,
{the localization} of the zero modes is as in Fig.~\ref{fermionwaveprofile_threeinterval}.
The concrete forms of mode functions for $\Psi$ are the same {as} in Eq.~(\ref{zeromodefunctionprofile2_twointerval}){,} and we know those for $\Psi'$ after adding {the} prime symbol ${}'$ {to} some parameters as {${\psi}_{1L}^{(0)} \rightarrow {{\psi'}_{1L}^{(0)}}, {\psi}_{2L}^{(0)} \rightarrow {{\psi'}_{2L}^{(0)}}, M_F \rightarrow M'_F, L_1 \rightarrow L'_1$ in Eq.~(\ref{zeromodefunctionprofile1_twointerval})}.

The remarkable point in the system with the two {point interactions} is that the two zero modes with different 4D chirality and ``generation" indices overlap in a middle region {($L_{1}<y<L_{1}'$)} of the system.
This means that flavor mixing {can be} realized in our configuration.
When we consider a UED-type scenario, {right- and left-handed} fermions are supplied by different 5D fields of $SU(2)_W$ doublet and singlet, respectively and separately.
Thereby{,} we can install this mechanism into {the} UED model with no obstacle.
In the SM, in fact, the flavor mixing structure is incorporated in the Yukawa sector with Yukawa couplings with the most general form and three copies of each $SU(2)_W$ doublet and singlet in the gauge eigenbases.
{These situations are} affected exceedingly after adapting our mechanism for realizing flavor mixing.
In addition, we should {consider} the profile of Higgs because it is a very important ingredient in {the} Yukawa sector.
{These issues will be discussed concretely in the next section.}
%
%
%

%%%%%%%%%%%%%%%%%%%%%%%%%%%%%%%%%%%%%%%%%%%%%
%%%%%%%%%%%%%%%%%%%%%%%%
\section{Searching for the possibility {of} achieving quark mass hierarchy and mixing in {a multiple} {point interaction} system
\label{assumedVEVanalysis_section}}
%%%%%%%%%%%%%%%%%%%%%%%%
%%%%%%%%%%%%%%%%%%%%%%%%%%%%%%%%%%%%%%%%%%%%%

Based on our discussion in Section~\ref{section:basic}{,}
we make an attempt to create a model where quark mass hierarchy and mixing are accomplished by {the} use of the mechanism of flavor mixing due to {multiple} {point interactions}.
In this section, we consider
the following form of action $S$ with effective Yukawa coupling in {the} new geometry:
\al{
S &= \int d^4 x \int_{0}^{L_3} dy \Bigg\{ \Big[ \overline{Q} \paren{i \pal_M \Gamma^M + M_{Q} } Q + \overline{\mathcal{U}} \paren{i \pal_M \Gamma^M + M_{\mathcal{U}} } \mathcal{U} \notag \\
&\phantom{=\int d^4 x \int_{0}^{L_3} \Bigg\{ \ \ \, } + \overline{\mathcal{D}} \paren{i \pal_M \Gamma^M + M_{\mathcal{D}} } \mathcal{D}
\Big] \notag  \\
&\phantom{=\int d^4 x \int_{0}^{L_3} \Bigg\{\ \  \, }  
- \Big[ Y^{(u)} \overline{U} {\langle \phi(y) \rangle} \mathcal{U}
+ Y^{(d)} \overline{D} {\langle \phi(y) \rangle} \mathcal{D}
+ \text{h.c.} \Big] \Bigg\}.
\label{effectiveaction} 
}
We have introduced one $SU(2)_W$ quark doublet $Q$, one up-type quark singlet $\mathcal{U}$, {and}
one down-type quark singlet $\mathcal{D}${,} with their 5D Dirac bulk masses $M_{Q}, M_{\mathcal{U}}, M_{\mathcal{D}}$.
{We would like to emphasize that our model does not possess
any generation index at this stage and that {fermion}
generation can appear dynamically.}
Here we have assumed that a ``Higgs doublet" $\mathcal{H}$ acquires VEV
with {$y$-position} dependence such as
\al{
\mathcal{H} = \begin{pmatrix} 0 \\ \langle \phi(y) \rangle \end{pmatrix}
}
and that the structure of the Yukawa sector is the same {as} that of the SM.
{Note that the} 5D up (down) quark Yukawa {couplings} $Y^{(u)}$ $(Y^{(d)})$ {also} do not contain {any} generation index {for} the quarks in our model.
The $SU(2)_W$ quark doublet can be decomposed as
\al{
Q = \begin{pmatrix} U \\ D\end{pmatrix}.
}

Before we go into more details, we would like to point out
some remarkable properties of our model.
The form of the action in Eq.~(\ref{effectiveaction}) seems to be very similar to
the corresponding part of the {mUED} model {at first glance, but} there are two significant differences {between the two theories, as explained below.}

One {concerns} the structure of {the} Yukawa coupling.
In the UED, we should introduce three copies of fermion configurations to realize {the} {three-generation} structure of the SM.
{The profiles of the} mode functions describing zero mode fermions take {constants, therefore} fine-tuning in the Yukawa couplings is inevitable.
On the other hand, we only introduce one copy of fermion configurations in our model.

\begin{figure}
\centering
\includegraphics[width=0.7\columnwidth]{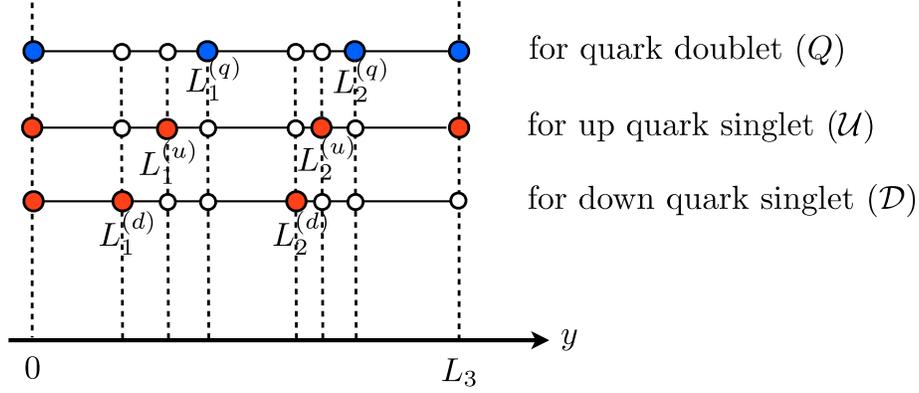}
\caption{
A schematic diagram for explaining our attempt.
{The conventions for the} circular spots and dots are the same {as} those in Fig.~\ref{overview_threeinterval}.
}
\label{systemprofile0_pdf}
\end{figure}

The other one can be seen in properties of the VEV of the Higgs boson.
In {mUED} models, BCs of the Higgs field are chosen as {the} Neumann-type,
then the VEV gets to be a constant, {which} is the same with the SM.
Here we consider the Yukawa structure in our model briefly.
As in Eq.~(\ref{effectiveaction}), the 5D Yukawa couplings do not possess any generation indices. 
The SM Yukawa structure is expected to be produced through geometry of an effective form of Higgs VEV and the lopsided wave functions of zero mode fermions.
In our model,  
the VEV profile of the Higgs scalar $\left\langle \phi \right\rangle$ is assumed to be $y$-position-dependent and to take the ``warped" form of
\al{
{\left\langle \phi(y) \right\rangle = \mathcal{A} \exp[ \alpha (y-L)]},
\label{assumedHiggsVEVprofile}
}
with two massive parameters $\mathcal{A}$ and $\alpha$, whose mass dimensions are $3/2$ and $1$, respectively.
{$L\,(=L_3)$} means the length of the total system.
The reason for forming this conjecture is {that} this shape of the VEV is preferable {for generating} the large quark mass hierarchy in the SM within a natural-ordered setting of parameters.
At this stage, we do not worry about the {precise method} of realizing this type of VEV profile{,} and mainly 
devote our attention to phenomenological issues of it.
In a later section, we discuss an example of {realizing} this (assumed) setup {using} the generalized Higgs BCs discussed in Ref.~\cite{Fujimoto:2011kf}.

We have to introduce new {point interactions}, {where we take the Dirichlet BC} for
all of $Q,\mathcal{U},\mathcal{D}$ to realize three generations in the SM.
{To this end, we assume that each 5D fermion feels nontrivially
two point interactions with the Dirichlet BC. It turns out
that {the} zero modes of the fermions are three-fold degenerate,
where each profile is confined in one of three segments.}
Here we consider the quark profiles explained in Fig.~\ref{systemprofile0_pdf}, where the meaning of the red and blue circular spots and the white dots are the same {as} those in Fig.~\ref{overview_threeinterval}.
It should be emphasized that {the} positions of {point interactions} which each 5D fermion feels are not necessarily common.
We assign the coordinates of the lower and upper ends of the total system {at} $0\,(=L_0)$ and $L_3$, respectively, and the locations of {the} {point interactions} between the two end points of the interval are represented by $L_1^{(q)}, L_2^{(q)}, L_1^{(u)}, L_2^{(u)}, L_1^{(d)}, L_2^{(d)}${,} with the superscripts {identifying} the type of the fields and the subscripts {showing} the sequences when we count them from the left to the right {(see Fig.~\ref{systemprofile0_pdf})}.

Concretely speaking, we adopt the following BCs for $Q, \U, \D$, respectively as
\al{
Q_R &= 0  \qquad \text{at} \quad y=0,\, L_1^{(q)} \pm \varepsilon,\, L_2^{(q)} \pm \varepsilon,\, L_3, \label{SMfermion_BC1} \\
\U_L &= 0  \qquad \text{at} \quad y=0,\, L_1^{(u)} \pm \varepsilon,\, L_2^{(u)} \pm \varepsilon,\, L_3, \label{SMfermion_BC2} \\
\D_L &= 0  \qquad \text{at} \quad y=0,\, L_1^{(d)} \pm \varepsilon,\, L_2^{(d)} \pm \varepsilon,\, L_3, \label{SMfermion_BC3}
}
where {the} three-fold generated {left- (right-)handed} zero modes emerge in $Q$ ($\U$, $\D$) as we have discussed before.
Here we {do not} describe the ``continuous" condition for each field.
It is noted that the profiles of $U$ and $D$ get to be the same in our configuration.
The orders of the positions are settled on as
\al{
0\,(=L_0) < L_1^{(u)} < L_1^{(q)} < L_2^{(u)} < L_2^{(q)} < L_3, \notag \\
0\,(=L_0) < L_1^{(d)} < L_1^{(q)} < L_2^{(d)} < L_2^{(q)} < L_3.
\label{positionorder1}
}

We {note that the} values of 4D effective Yukawa masses are evaluated as those of overlap integrals among the VEV, {right- and left-handed} modes. Then both magnitudes and signs of the bulk masses $M_Q, M_{\mathcal{U}}, M_{\mathcal{D}}$ govern an important part of the results.

The forms of {the} effective 4D Yukawa {masses} among the three generations are obtained after integration {over} $y$ as
\al{
-\int_0^{L_3} dy \Big[ Y^{(u)} \overline{U} \langle \phi(y) \rangle \mathcal{U} +
Y^{(d)} \overline{D} \langle {\phi(y)} \rangle \mathcal{D} + \text{h.c.} \Big]
&= -\begin{bmatrix} \overline{u}^{(0)}_{1L}(x), \overline{u}^{(0)}_{2L}(x), \overline{u}^{(0)}_{3L}(x) \end{bmatrix}
{\mathcal{M}^{(u)}}
\begin{bmatrix}
u^{(0)}_{1R} \\
u^{(0)}_{2R} \\
u^{(0)}_{3R}
\end{bmatrix} \notag \\
&\phantom{=\ }
-\begin{bmatrix} \overline{d}^{(0)}_{1L}(x), \overline{d}^{(0)}_{2L}(x), \overline{d}^{(0)}_{3L}(x) \end{bmatrix}
{\mathcal{M}^{(d)}}
\begin{bmatrix}
d^{(0)}_{1R} \\
d^{(0)}_{2R} \\
d^{(0)}_{3R}
\end{bmatrix}
+\text{h.c.},
}
where the mass matrices {${\mathcal{M}}^{(u)}$ and ${\mathcal{M}}^{(d)}$} have the following structure{:}
\al{
\mathcal{M}^{(u)} =
\begin{bmatrix}
m^{(u)}_{11} & m^{(u)}_{12} & 0 \\
0 & m^{(u)}_{22} & m^{(u)}_{21} \\
0 & 0 & m^{(u)}_{33}  
\end{bmatrix}, \quad
\mathcal{M}^{(d)} =
\begin{bmatrix}
m^{(d)}_{11} & m^{(d)}_{12} & 0 \\
0 & m^{(d)}_{22} & m^{(d)}_{21} \\
0 & 0 & m^{(d)}_{33}  
\end{bmatrix}.
\label{massmatrix1}
}
The three-fold degenerated zero modes are distinguished by the generation indices ``$1,2,3$" in both up-type {$(u)$ and down-type $(d)$ quarks}.
Each matrix component in Eq.~(\ref{massmatrix1}) is calculated by overlap integrals among the Higgs VEV and zero mode functions of {right- and left-handed} fermions, which are functions of the fermion bulk masses $M_Q, M_{\mathcal{U}}, M_{\mathcal{D}}$ and the locations of the {point interactions} $L_1^{(q)}, L_2^{(q)}, L_1^{(u)}, L_2^{(u)}, L_1^{(d)}, L_2^{(d)}$.
{In contrast to} the SM, some elements of the Yukawa mass matrices are zero.
The diagonal parts $m^{(u)}_{11}, m^{(u)}_{22}, m^{(u)}_{33}, m^{(d)}_{11}, m^{(d)}_{22}, m^{(d)}_{33}$ are always nonzero unless we take an extremal parameter choice, {e.g.}, $L_1^{(u)} = L_2^{(u)}$, which, of course, is unsuitable for our purpose in this paper.
{We notice that the characteristic form of the mass matrices in Eq.~(\ref{massmatrix1}) is given by the geometry of our setting. Some of the components in the mass matrices vanish due to no overlapping of mode functions.}

Which non-diagonal component is nonzero depends on {the positions} of the {point interactions}.
Following the rule in Eq.~(\ref{positionorder1}), {only the} $(1,2)$ and $(2,3)$ elements are nonzero.
When we reverse the order of position between $L_{1}^{(q)}$ and $L_{1}^{(u)}$ as
\al{
L_{1}^{(u)} < L_{1}^{(q)} \rightarrow L_{1}^{(q)} < L_{1}^{(u)},
}
the value of the $(1,2)$ component {vanishes $(m^{(u)}_{12} = 0)$} but that of the $(2,1)$ component becomes nonzero {$(m^{(u)}_{21} \neq 0)$}.
This issue is easily understandable when we focus on the fact that the row (column) index of the mass matrix corresponds to the generation of the {left- (right-) handed} fermion, respectively.
{We would like to mention that} in our model, flavor mixing is inevitable to realize the SM quark masses.
Because in the aligned configuration of $L_1^{(q)} = L_1^{(u)} = L_1^{(d)},\, L_2^{(q)} = L_2^{(u)} = L_2^{(d)}$, it is very hard to realize the quark mass patterns of the first generation ($m_{\text{up}} < m_{\text{down}}$) and those of the third generation ($m_{\text{top}} > m_{\text{bottom}}$) simultaneously.

The point is whether {or not} we can reproduce the structure of the CKM matrix based on the limited forms of the Yukawa mass matrices, where we never find {overlap between the} first and third generations.
{After some calculations, we can find a set of input parameters where we reproduce the quark mass hierarchy and the CKM matrix simultaneously.
{In the next subsection we discuss, in detail, whether we can
simultaneously reproduce the structure of the quark mass hierarchy
and the CKM matrix based on the limited forms of the Yukawa mass
matrices (\ref{massmatrix1}) without {overlapping} between the first and third
generations.}

%%%%%%%%%%%%%%%%%%%%%%%%%%%%%%%%%%%%%%%%%%%%%
\subsection{{A solution in the setup with a warped VEV}}
%%%%%%%%%%%%%%%%%%%%%%%%%%%%%%%%%%%%%%%%%%%%%

Let us first give the concrete forms of the Yukawa mass matrices in {our system} which {were} discussed in the previous subsection.
The fields $Q, \mathcal{U}, \mathcal{D}$  with the BCs in Eqs~(\ref{SMfermion_BC1}), (\ref{SMfermion_BC2}){,} and (\ref{SMfermion_BC3}) are KK-decomposed as follows:
\al{
Q(x,y) = \begin{pmatrix} U(x,y) \\ D(x,y) \end{pmatrix} &=
\begin{pmatrix}
\sum_{i=1}^{3} u^{(0)}_{iL}(x) f_{q^{(0)}_{iL}}(y) \\
\sum_{i=1}^{3} d^{(0)}_{iL}(x) f_{q^{(0)}_{iL}}(y)
\end{pmatrix}
+ (\text{KK modes}), \\
\mathcal{U}(x,y) &=
\sum_{i=1}^{3} u^{(0)}_{iR}(x) f_{u^{(0)}_{iR}}(y) + (\text{KK modes}), \\
\mathcal{D}(x,y) &=
\sum_{i=1}^{3} d^{(0)}_{iR}(x) f_{d^{(0)}_{iR}}(y) + (\text{KK modes}),
}
where we only focus on the zero mode parts.
Here{,} the zero mode functions are obtained in the following forms{:}
\al{
f_{q^{(0)}_{iL}}(y) &= 
		\mathcal{N}_{i}^{(q)} e^{M_{Q} (y-L_{i-1}^{(q)})} \Big[ \theta(y-L_{i-1}^{(q)}) \theta(L_{i}^{(q)}-y) \Big], \label{doubletwavefunction} \\
f_{u^{(0)}_{iR}}(y) &= 
		\mathcal{N}_{i}^{(u)} e^{-M_{\mathcal{U}} (y-L_{i-1}^{(u)})} \Big[ \theta(y-L_{i-1}^{(u)}) \theta(L_{i}^{(u)}-y) \Big], \label{upsingletwavefunction} \\
f_{d^{(0)}_{iR}}(y) &= 
		\mathcal{N}_{i}^{(d)} e^{-M_{\mathcal{D}} (y-L_{i-1}^{(d)})} \Big[ \theta(y-L_{i-1}^{(d)}) \theta(L_{i}^{(d)}-y) \Big], \label{downsingletwavefunction}
}
where we use the conventions, for clarity, of
\al{
\Delta L^{(l)}_{i} &= L^{(l)}_{i} - L^{(l)}_{i-1} \qquad
(\text{for } i=1,2,3;\ l=q,u,d), \notag \\
0\,(=L_0) &= L_0^{(q)} =L_0^{(u)} {=L_0^{(d)}}, \notag \\
L_3 &= L_3^{(q)} =L_3^{(u)} {=L_3^{(d)}}, \notag
}
\vspace{-8mm}
\al{
\mathcal{N}_{i}^{(q)} = \sqrt{\frac{2M_{Q}}{e^{2M_{Q} \Delta L_{i}^{(q)} }-1}}, \quad
\mathcal{N}_{i}^{(u)} = \sqrt{\frac{2M_{\mathcal{U}}}{1-e^{-2M_{\mathcal{U}} \Delta L_{i}^{(u)} }}}, \quad
\mathcal{N}_{i}^{(d)} = \sqrt{\frac{2M_{\mathcal{D}}}{1-e^{-2M_{\mathcal{D}} \Delta L_{i}^{(d)} }}}.
\label{wavefunctionconventions}
}
$\mathcal{N}_{i}^{(q)}, \mathcal{N}_{i}^{(u)}, \mathcal{N}_{i}^{(d)}$ are wavefunction normalization factors for $f_{q^{(0)}_{iL}}, f_{u^{(0)}_{iL}}, f_{d^{(0)}_{iL}}$, respectively.
The length of the total system $L$ takes the universal value of
\al{
L := L_3 = L^{(q)}_{3} - L^{(q)}_{0} = L^{(u)}_{3} - L^{(u)}_{0} = L^{(d)}_{3} - L^{(d)}_{0}.
\label{totallength}
}
\begin{figure}
\centering
\includegraphics[width=0.8\columnwidth]{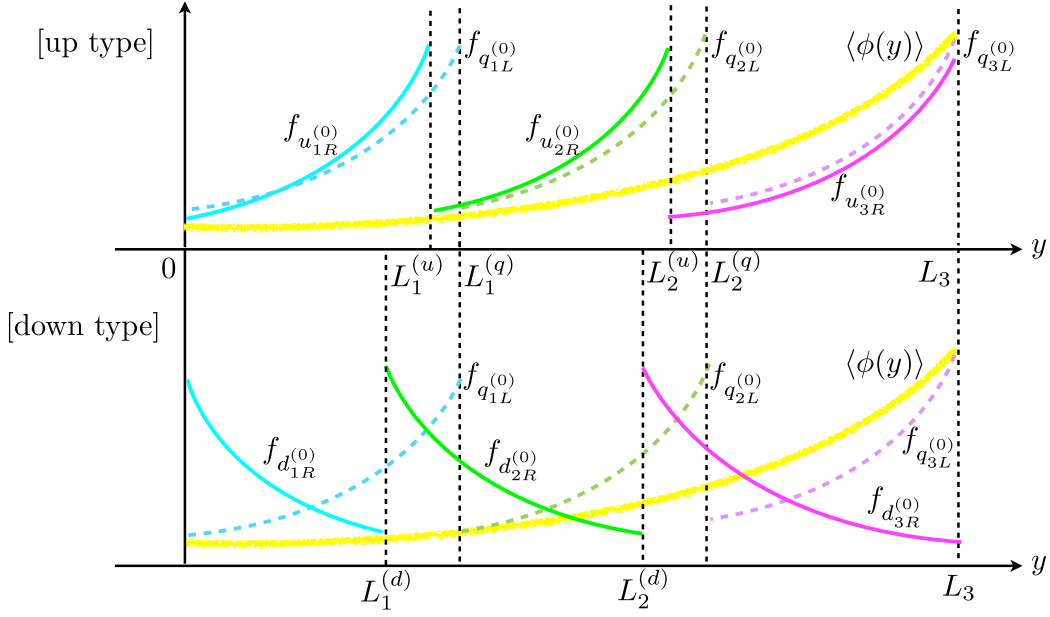}
\caption{
An {outline} of the {wavefunction} profiles.
}
\label{overlapprofile_pdf}
\end{figure}
{We} choose the signs of the fermion bulk masses $M_Q, M_\U, M_\D$ as
\al{
M_Q > 0, M_\U < 0, M_\D > 0,
}
where we take a negative value in $M_\U$ to generate a large {overlap} for the top quark mass (mainly) in $m_{33}^{(u)}$.
An {outline} of the wavefunction profiles {is shown} in Fig.~\ref{overlapprofile_pdf}.

Each component of  {$\mathcal{M}^{(u)}$ and $\mathcal{M}^{(d)}$} is acquired by calculating the corresponding overlap integral as follows:
\al{
m^{(u)}_{11} &= {Y}^{(u)} \int_{0}^{L_{1}^{(u)}} \!\! dy 
		f_{q^{(0)}_{1L}}(y) f_{u^{(0)}_{1R}}(y) \langle \phi(y) \rangle \notag \\
&= \mathcal{N}_{1}^{(q)} \mathcal{N}_{1}^{(u)} {Y}^{(u)} \mathcal{A}{e^{-\alpha L} }
		\Bigg\{ \frac{e^{(M_{Q} + |M_{\U}| + \alpha)L_{1}^{(u)}}-1}
		{M_{Q} + |M_{\U}| + \alpha} \Bigg\}, \label{Mu11}\\
m^{(u)}_{22} &= {Y}^{(u)} \int_{L_1^{(q)}}^{L_{2}^{(u)}} \!\! dy 
		f_{q^{(0)}_{2L}}(y) f_{u^{(0)}_{2R}}(y) \langle \phi(y) \rangle \notag \\
&= \mathcal{N}_{2}^{(q)} \mathcal{N}_{2}^{(u)} {Y}^{(u)} \mathcal{A} {e^{-\alpha L} }
		\Bigg\{ \frac{e^{(M_{Q} + |M_{\U}| + \alpha)L_{2}^{(u)}}-e^{(M_{Q} + |M_{\U}| + \alpha)L_{1}^{(q)}}}
		{M_{Q} + |M_{\U}| + \alpha} \Bigg\}
		e^{-M_{Q} L_{1}^{(q)} -|M_{\U}| L_1^{(u)}}, \\
m^{(u)}_{33} &= {Y}^{(u)} \int_{L_2^{(q)}}^{L_{3}^{(q)}} \!\! dy 
		f_{q^{(0)}_{3L}}(y) f_{u^{(0)}_{3R}}(y) \langle \phi(y) \rangle \notag \\
&= \mathcal{N}_{3}^{(q)} \mathcal{N}_{3}^{(u)} {Y}^{(u)} \mathcal{A} {e^{-\alpha L} }
		\Bigg\{ \frac{e^{(M_{Q} + |M_{\U}| + \alpha)L_{3}^{(q)}}-e^{(M_{Q} + |M_{\U}| + \alpha)L_{2}^{(q)}}}
		{M_{Q} + |M_{\U}| + \alpha} \Bigg\}
		e^{-M_{Q} L_{2}^{(q)} -|M_{\U}| L_{2}^{(u)}},}
\al{
m^{(u)}_{12} &= {Y}^{(u)} \int_{L_1^{(u)}}^{L_{1}^{(q)}} \!\! dy 
		f_{q^{(0)}_{1L}}(y) f_{u^{(0)}_{2R}}(y) \langle \phi(y) \rangle \notag \\
&= \mathcal{N}_{1}^{(q)} \mathcal{N}_{2}^{(u)} {Y}^{(u)} \mathcal{A} {e^{-\alpha L} }
		\Bigg\{ \frac{e^{(M_{Q} + |M_{\U}| + \alpha)L_{1}^{(q)}}-e^{(M_{Q} + |M_{\U}| + \alpha)L_{1}^{(u)}}}
		{M_{Q} + |M_{\U}| + \alpha} \Bigg\}
		e^{-|M_{\U}| L_{1}^{(u)}}, \\
m^{(u)}_{23} &= {Y}^{(u)} \int_{L_2^{(u)}}^{L_{2}^{(q)}} \!\! dy 
		f_{q^{(0)}_{2L}}(y) f_{u^{(0)}_{3R}}(y) \langle \phi(y) \rangle \notag \\
&= \mathcal{N}_{2}^{(q)} \mathcal{N}_{3}^{(u)} {Y}^{(u)} \mathcal{A} {e^{-\alpha L} }
		\Bigg\{ \frac{e^{(M_{Q} + |M_{\U}| + \alpha)L_{2}^{(q)}}-e^{(M_{Q} + |M_{\U}| + \alpha)L_{2}^{(u)}}}
		{M_{Q} + |M_{\U}| + \alpha} \Bigg\}
		e^{-M_{Q} L_{1}^{(q)} -|M_{\U}| L_{2}^{(u)}},\\
m^{(d)}_{11} &= {Y}^{(d)} \int_{0}^{L_{1}^{(d)}} \!\! dy 
		f_{q^{(0)}_{1L}}(y) f_{d^{(0)}_{1R}}(y) \langle \phi(y) \rangle \notag \\
&= \mathcal{N}_{1}^{(q)} \mathcal{N}_{1}^{(d)} {Y}^{(d)} \mathcal{A} {e^{-\alpha L} }
		{\Bigg\{ \frac{e^{(M_{Q} - M_{\D} + \alpha)L_{1}^{(d)}} - 1}
		{M_{Q} - M_{\D} + \alpha} \Bigg\}}, \\
m^{(d)}_{22} &= {Y}^{(d)} \int_{L_1^{(q)}}^{L_{2}^{(d)}} \!\! dy 
		f_{q^{(0)}_{2L}}(y) f_{d^{(0)}_{2R}}(y) \langle \phi(y) \rangle \notag \\
&= \mathcal{N}_{2}^{(q)} \mathcal{N}_{2}^{(d)} {Y}^{(d)} \mathcal{A} {e^{-\alpha L} }
		\Bigg\{ \frac{e^{(M_{Q} - M_{\D} + \alpha)L_{2}^{(d)}}-e^{(M_{Q} - M_{\D} + \alpha)L_{1}^{(q)}}}
		{M_{Q} - M_{\D} + \alpha} \Bigg\}
		e^{-M_Q L_1^{(q)}  + M_{\D} L_{1}^{(d)}}, \\
m^{(d)}_{33} &= {Y}^{(d)} \int_{L_2^{(q)}}^{L_{3}^{(q)}} \!\! dy 
		f_{q^{(0)}_{3L}}(y) f_{d^{(0)}_{3R}}(y) \langle \phi(y) \rangle \notag \\
&= \mathcal{N}_{3}^{(q)} \mathcal{N}_{3}^{(d)} {Y}^{(d)} \mathcal{A} {e^{-\alpha L} }
		\Bigg\{ \frac{e^{(M_{Q} - M_{\D} + \alpha)L_{3}^{(q)}}-e^{(M_{Q} - M_{\D} + \alpha)L_{2}^{(q)}}}
		{M_{Q} - M_{\D} + \alpha} \Bigg\}
		{e^{-M_{Q}L^{(q)}_{2}+M_{\mathcal{D}} L^{(d)}_{2}}},\\
m^{(d)}_{12} &= {Y}^{(d)} \int_{L_1^{(d)}}^{L_{1}^{(q)}} \!\! dy 
		f_{q^{(0)}_{1L}}(y) f_{d^{(0)}_{2R}}(y) \langle \phi(y) \rangle \notag \\
&= \mathcal{N}_{1}^{(q)} \mathcal{N}_{2}^{(d)} {Y}^{(d)} \mathcal{A} {e^{-\alpha L} }
		\Bigg\{ \frac{e^{(M_{Q} - M_{\D} + \alpha)L_{1}^{(q)}}-e^{(M_{Q} - M_{\D} + \alpha)L_{1}^{(d)}}}
		{M_{Q} - M_{\D} + \alpha} \Bigg\}
		e^{M_{\D} L_{1}^{(d)}}, \\
m^{(d)}_{23} &= {Y}^{(d)} \int_{L_2^{(d)}}^{L_{2}^{(q)}} \!\! dy 
		f_{q^{(0)}_{2L}}(y) f_{d^{(0)}_{3R}}(y) \langle \phi(y) \rangle \notag \\
&= \mathcal{N}_{2}^{(q)} \mathcal{N}_{3}^{(d)} {Y}^{(d)} \mathcal{A} {e^{-\alpha L} }
		\Bigg\{ \frac{e^{(M_{Q} - M_{\D} + \alpha)L_{2}^{(q)}}-e^{(M_{Q} - M_{\D} + \alpha)L_{2}^{(d)}}}
		{M_{Q} - M_{\D} + \alpha} \Bigg\}
		e^{-M_Q L_1^{(q)} + M_{\D} L_{2}^{(d)}}, \label{Md13}
}
where we {have used} the forms of {the} wavefunctions in Eqs.~(\ref{doubletwavefunction}), (\ref{upsingletwavefunction}), (\ref{downsingletwavefunction}), the conventions in Eq.~(\ref{wavefunctionconventions}), and the assumed profiles of the Higgs VEV in Eq.~(\ref{assumedHiggsVEVprofile}).
In this paper, we choose the parameters as
\beq
\begin{array}{lll}
L_{1}^{(q)} = 0.338 \cdot L, & L_{2}^{(q)} = 0.689 \cdot L, &
		L_{3}^{(q)} = 1 \cdot L,  \\
L_{1}^{(u)} = 0.0115 \cdot L, & L_{2}^{(u)} = 0.540 \cdot L, &
		L_{3}^{(u)} = 1 \cdot L,  \\
		L_{1}^{(d)} = 0.223 \cdot L, & L_{2}^{(d)} = 0.676 \cdot L, &
		L_{3}^{(d)} = 1 \cdot L, \\
M_Q = 6.67 \cdot L^{-1}, & M_\U = - 7.98 \cdot L^{-1}, & M_\D = 6.16 \cdot L^{-1}, \\
\alpha = 8.67 \cdot L^{-1}, & Y^{(u)}/Y^{(d)} = 12.0, & \mathcal{A} Y^{(u)} = 275\,\text{GeV},
\end{array}
\label{a_setofsolution2}
\eeq
which {reproduce} the quark mass hierarchy and the CKM matrix with {good precision}.
The quark mass eigenvalues and the CKM matrix are evaluated from Eqs.(\ref{Mu11})--(\ref{Md13}).
The diagonalized Yukawa mass matrices take the {forms}
{
\al{
\mathcal{M}^{(u)}|_{\text{diagonal}} &= \text{diag}\, (2.13\,\text{MeV},\ 1.18\,\text{GeV},\ 174\,\text{GeV}), \label{obtainedquarkmass1} \\
\mathcal{M}^{(d)}|_{\text{diagonal}} &= \text{diag}\, (3.85\,\text{MeV},\ 110\,\text{MeV},\ 4.19\,\text{GeV}),
\label{obtainedquarkmass2}
}
and the CKM matrix is given as
\al{
|V_{\text{CKM}}|=
\begin{bmatrix}
		0.976 & 0.213 &  0.00448 \\
		0.213 & 0.976 & 0.0475 \\
		0.0145 & 0.0454 & 0.999
\end{bmatrix}.
\label{obtainedCKMmatrix}
}
}

\begin{table}[t]
\begin{center}
\begin{tabular}{|c|c||c|c|}
\hline 
up quark & mass & down quark & mass \\
\hline 
up\,$(u)$ &{$1.8$--$3.0$\,MeV} & down\,$(d)$ & {$4.5$--$5.5$\,MeV}\\
charm\,$( c )$ & {$1.250$--$1.300$\,GeV} & strange\,$(s)$ & {$90$--$100$\,MeV} \\
top\,$(t)$ & {$172.1$--$174.9$\,GeV} & bottom\,$(b)$ & {$4.15$--$4.21$\,GeV} \\
\hline
\end{tabular}
\caption{{Experimental values of quark masses from Ref.~\cite{Beringer:1900zz}.}}
\label{SMquarkmassinputs}
\end{center}
\end{table}
{
{Now}, we present the {latest} experimental values. The quark masses are summarized in Table~\ref{SMquarkmassinputs} and the CKM matrix is
\al{
|V_{\text{CKM}}|_{\text{exp.}}  =
\begin{bmatrix}
		0.974 & 0.225 & 0.00415 \\
		0.230 & 1.006 & 0.0409 \\
		0.0084 & 0.0429 & 0.89
\end{bmatrix}
}
from Ref.~\cite{Beringer:1900zz}.
}
From the results in Eqs.~(\ref{obtainedquarkmass1}) and (\ref{obtainedquarkmass2}), the mass eigenvalues are well reproduced within about {twenty percent,} and the {absolute values of the CKM matrix elements are} also described within about {twenty percent} in almost all the elements.
{But we should comment on the $(3,1)$ component of the CKM matrix which we have obtained, where
the deviation ratio measures up to $\sim 70 \%$.}

At this stage, we ponder over the number of {input} parameters and constraints {originating} from the equations.
After keeping the relations {concerning} the positions in Eqs.~(\ref{wavefunctionconventions}) and (\ref{totallength}) in mind, we count the independent d.o.f. (degrees of freedom) as
$7$ {for the positions} of {the} boundary points, $3$ for {the} fermion bulk masses, $2$ for {the} 5D Yukawa couplings, $2$ for {the} effective Higgs VEV parameters, respectively, and the total d.o.f. is $14$.
Meanwhile, the total number of {constraints} is {$9$, {consisting of} 6 quark mass eigenvalues and 3 CKM mixing angles.
Here the} two massive parameters of
\al{
Y^{(u)}\, (\text{or } Y^{(d)},\, \mathcal{A}),\  L
}
are still not specified.
It is noted that once one of $Y^{(u)}, Y^{(d)}, \mathcal{A}$ is fixed, {the others are} determined simultaneously.
The reason for this indetermination is that the electroweak scale is not indicated within our analysis.

The results in Eq.~(\ref{a_setofsolution2}) are realized broadly within $\mathcal{O}(10)$ range (when we choose the basis of scaling as $L^{-1}$) and
it is shown that we can explain {both} the quark mass hierarchy and the structure of the CKM matrix in a natural statement. {}

%%%%%%%%%%%%%%%%%%%%%%%%%%%%%%%%%%%%%%%%%%%%%
%%%%%%%%%%%%%%%%%%%%%%%%
\section{An example for realizing {{warped}} Higgs VEV by generalized Higgs boundary condition
\label{concretemodel_section}}
%%%%%%%%%%%%%%%%%%%%%%%%
%%%%%%%%%%%%%%%%%%%%%%%%%%%%%%%%%%%%%%%%%%%%%

In the previous section, we have verified the possibility of {obtaining} the quark mass hierarchy and the structure of the CKM matrix at the same instant without fine-tuning of {the} parameters in the setup, whose geometry contains many {point interactions}.
Here we have assumed the forms of 5D Yukawa couplings and the {warped} shape of the Higgs VEV profile $\langle \phi(y) \rangle$.
In this section, we give an example of generating this situation without conflicting with the physics {of} the SM.

%%%%%%%%%%%%%%%%%%%%%%%%%%%%%%%%%%%%%%%%%%%%%
\subsection{Generalized boundary condition for {a scalar}}
%%%%%%%%%%%%%%%%%%%%%%%%%%%%%%%%%%%%%%%%%%%%%

Following~\cite{Fujimoto:2011kf},
we can consider the physics which {is} described by the 5D actions on a single interval $[0,L]$ for a {5D scalar} $\Phi$ of
\al{
{S_{\Phi}} &= \int d^4 x \int_{0}^{L} dy \Big\{
		\Phi^{\dagger} \pal^{\mu} \pal_{\mu} \Phi
		+ \Phi^{\dagger} \pal_{y}^2 \Phi
		- V(|\Phi|^2) \Big\},
}
with the bulk potential $V(|\Phi|^2) = M^2 |\Phi|^2 + \frac{\lambda}{2} |\Phi|^4${,} and where we do not consider {a tree-level} brane-localized term.
The parameters {$M^2$, $\lambda$} need to be real due to hermiticity.
We can find a difference from the ordinary UED-type form in the structure of the derivatives.
Based on {the} variational principle, the following form should vanish at the boundaries $y=0, L$:
\al{
{\Phi^{\dagger} \pal_y \delta \Phi -(\pal_y \Phi)^{\dagger} \delta \Phi = 0 
		\qquad \text{at} \quad y=0, L.}
}
It turns out that a larger class of BCs is allowed with two new real massive parameters $L_{\pm}$, whose mass dimensions are $-1$, as
\al{
\Phi + L_+ \pal_y \Phi &= 0 \qquad \text{at} \quad y=0, \notag \\
\Phi - L_- \pal_y \Phi &= 0 \qquad \text{at} \quad y=L,
\label{RobinBC}
}
where $L_{\pm}$ can take values in the range of $-\infty \leq L_{\pm} \leq \infty$.
This type of BC is called {a} Robin boundary condition.
It is obvious that the above conditions include {the} ordinary Neumann (Dirichlet) BC in the case of ${L_{\pm}}=\pm \infty$ (${L_{\pm}}=0$).

%%%%%%%%%%%%%%%%%%%%%%%%%%%%%%%%%%%%%%%%%%%%%
\subsection{Position-dependent {scalar} VEV {with} generalized boundary condition
\label{Position-dependent Higgs VEV in generalized boundary condition}}
%%%%%%%%%%%%%%%%%%%%%%%%%%%%%%%%%%%%%%%%%%%%%

In this subsection, we discuss the profile of the scalar singlet $\Phi$ and its phase structure.
In the mUED model, where the Higgs doublet takes the Neumann BC at {both} boundaries, the 4D effective Higgs potential is minimized with ease, where the VEV is a constant ($y$-independent){,} and the situation is the same with the SM.

On the other hand in our model, the phase structure is nontrivial because the massive parameters $L_{\pm}$
emerge in our generalized BC in Eq.~(\ref{RobinBC}).
To investigate {the properties of the} phase structure of the {scalar} $\Phi$, we have to solve the minimization problem of the functional indicating 4D effective Higgs potential 
\al{
\E[\Phi] := \int_0^{L} dy \bigg\{ - \Phi^{\dagger} \pal_y^2 \Phi + V(|\Phi|^2) \bigg\}
= \int_0^{L} dy \bigg\{ - \Phi^{\dagger} \pal_y^2 \Phi + M^2 |\Phi|^2 + \frac{\lambda}{2} |\Phi|^4 \bigg\}.
\label{Higgseffectivepotential}
}
It is important {to} incorporate the kinetic term along {with} the extra {spatial} direction into the potential because the minimum configuration of the singlet $\Phi$ probably possesses
$y$-dependence, as we will see below.

Hereafter{,} we search for the form of the VEV $\langle \Phi(y) \rangle$ by solving the EOM
\al{
(-\pal_y^2 + {M^2})\Phi + \lambda \Phi^{\dagger} \Phi^2 = 0,
\label{HiggsEOM}
}
which can be obtained after taking variation in Eq.~(\ref{Higgseffectivepotential}).
The solutions of Eq.~(\ref{HiggsEOM}) are {generally} found to be expressed in terms of
Jacobi's elliptic functions. The types of {solution} with
$M^{2}>0$ are classified based on a parameter $Q$, which is an
integration constant after integrating along $y$, with mass
dimensions 5. It turns out that with the choice of $Q<0${,} a solution of Eq.~(\ref{HiggsEOM}), which can realize {the desired
``warped"} VEV in an asymptotic form, is {given} by
\al{
\vev{\Phi(y)} = {\nu} \cdot \frac{1}{\text{cn}\paren{\sqrt{\frac{\lambda}{2}} \frac{\mu}{k} (y-y_0), k}},
}
where the several parameters are defined as
\al{
\mu^2 &= \frac{M^2}{\lambda} \paren{1 + \sqrt{1 + \frac{4 \lambda |Q|}{M^4}} },
\label{form_mu} \\
\nu^2 &= \frac{M^2}{\lambda} \paren{\sqrt{1 + \frac{4 \lambda |Q|}{M^4}}-1},
\label{form_nu} \\
k^2 &= \frac{\mu^2}{\mu^2 + \nu^2}.
}
$y_0$ means a translation d.o.f. along $y$, which appears after integration {to solve} the equation in~(\ref{HiggsEOM}).
The index $k$ is an important parameter of elliptic function {for determining} the profile.
We can rewrite $k$ by use of the input parameters in $\mu,\nu$ as
\al{
k = { \sqrt{\frac{1}{2}\paren{1 + \frac{1}{\sqrt{1+X}}}} } \quad \paren{X:= \frac{4\lambda|Q|}{M^4}},
\label{parameter_k}
}
where we conclude that the possible region of the value of $k$ is
\al{
\sqrt{\frac{1}{2}} \leq k \leq 1.
}
Here we {note} that the condition $\lambda > 0$ is required to {ensure} the stability of the vacuum.
We again write down the form of $\Phi$ with less abbreviation:
\al{
\vev{\Phi(y)} = \sqbr{ \frac{M}{\sqrt{\lambda}} \br{\sqrt{1 + X}-1}^{1/2} }
\times  \frac{1}{ \text{cn}  \paren{ M \br{1 + X}^{1/4} (y-y_0), \sqrt{\frac{1}{2}\paren{1 + \frac{1}{\sqrt{1+X}}}}       }  }.
\label{exactHiggsVEVform}
}

Next{,} we decide to choose our strategy to treat {the} scalar BC.
We rewrite Eq.~(\ref{RobinBC}) as follows:
\al{
L_+ = - \vev{\Phi(0)}/\pal_y \vev{\Phi(0)}, \qquad L_- = + \vev{\Phi(L)}/\pal_y \vev{\Phi(L)}.
\label{inverseLs}
}
The {$L_+$, $L_-$} are determined by the values of {$\vev{\Phi(y)}$, $\pal_y \vev{\Phi(y)}$} at the endpoints $y=0,L${, and we first} discuss the shapes of {$\vev{\Phi(y)}$, $\pal_y \vev{\Phi(y)}$}.
{In Fig.~\ref{VEVplots}, we show three plots of {$\vev{\Phi(y)}$, $\pal_y \vev{\Phi(y)}$} and ${\vev{\Phi(y)}}/{\pal_y \vev{\Phi(y)}}$, with suitable normalizations.}
The red, blue, magenta, green curves are in $k=0,0.71,0.9,1$, respectively and
the red dots are the points $y=\pi/2$.
{Some comments are in order:}
\begin{figure}
\centering
%\hspace{-12mm}
\includegraphics[width=\columnwidth]{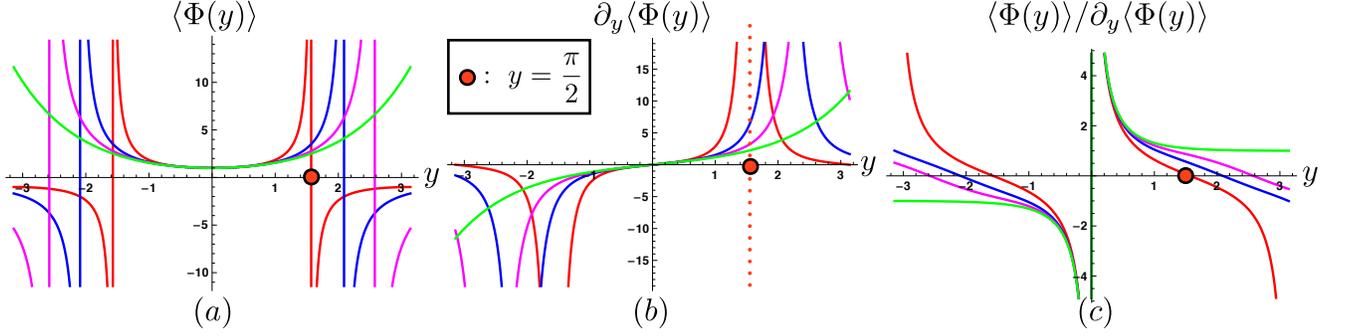}
\caption{
{These three plots represent the VEV profile of $(a)$ $\vev{\Phi(y)} = 1/\text{cn}\paren{y,k}$, $(b)$ its first derivative $\pal_y \vev{\Phi(y)}$,
and $(c)$ the form of $\vev{\Phi(y)}/\pal_y \vev{\Phi(y)}$.}
The red, blue, magenta, green curves are in $k=0,0.71,0.9,1$, respectively.
The red dots are the points $y=\pi/2$.
\label{VEVplots_pdf}
}
\label{VEVplots}
\end{figure}
\begin{itemize}
\item The function of $1/\text{cn}(y)$ can be represented as {a} trigonometric or hyperbolic {functions} in extremal cases of $k$, {namely} $1/\text{cn}(y)|_{k=0} = 1/\cos(y)$ and $1/\text{cn}(y)|_{k=1} = \cosh(y)$.
Following the change of the value of $k$ from $0$ to $1$, the profile of $1/\text{cn}(y)$ smoothly {shifts} from $1/\cos(y)$ to $\cosh(y)$.
\item $\vev{\Phi(y)}|_{k=0}$ and $\pal_y \vev{\Phi(y)}|_{k=0}$ are divergent at $y=\pi/2$. Increasing the value of $k$ from $0$ to $1$,
this divergent point moves from $\pi/2$ to infinity.
\item The profile of $[\vev{\Phi(y)}/\pal_y \vev{\Phi(y)}]|_{k=0}$ is divergent at $y=\pi$ and takes zero value at $y=\pi/2$.
As the value of $k$ {increases} from $0$ to $1$,
these points move from $\pi \,(\pi/2)$ to infinity.
This profile is also divergent at $y=0$ independently of the value of $k$.
\item In the region between $y=0$ and $y=y_d$, where $y_d$ is {the} first divergent point with positive value  and, of course, $y_d = \pi/2$ in {the} case of $k=0${, the} profile of $\vev{\Phi(y)}/\pal_y \vev{\Phi(y)}$ is monotonically decreasing independently of the value of $k$.
\end{itemize}
In this paper, we focus on the segment of ${(0,y_d)}$.
{For} this segment, as we review above, the profile of $\vev{\Phi(y)}/\pal_y \vev{\Phi(y)}$ is monotonically decreasing independently of the value of $k$ and {takes} positive value.
Therefore, the values of $L_+,L_-$ obey the following condition:
\beq
\left\{
\begin{array}{l}
L_+ < 0 \\
L_- > 0 \\
|L_+| > |L_-|
\end{array},
\right.
\label{L_condition}
\eeq
where the condition {$L_+ + L_- < 0$} is automatically derived from the conditions of Eq.~(\ref{L_condition}).
\begin{figure}
\centering
%\hspace{-12mm}
\includegraphics[width=0.5\columnwidth]{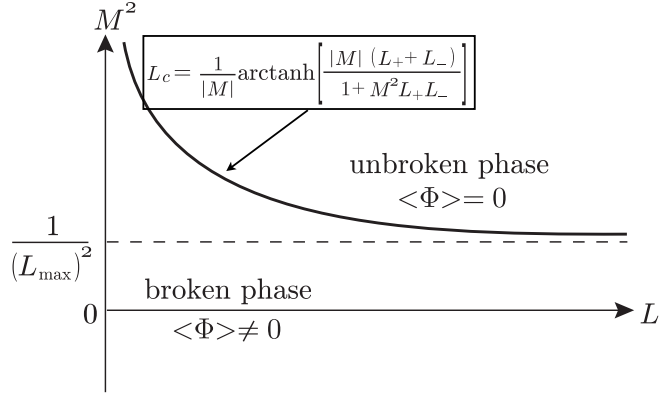}
\caption{
Phase diagram of the {scalar} $\Phi$ in the case of $L_+ L_- < 0, L_+ + L_- \leq 0$ {from} Ref.~\cite{Fujimoto:2011kf}.
}
\label{Higgsphasediagram_pdf}
\end{figure}
The phase structure of the {scalar} singlet $\Phi$ with generalized BC is explored in Ref.~\cite{Fujimoto:2011kf} and we quote a phase diagram in our case of $L_+ L_- < 0, L_+ + L_- \leq 0$  as Fig.~\ref{Higgsphasediagram_pdf}.
$L_{\text{max}}$ in Fig.~\ref{Higgsphasediagram_pdf} is defined as
\al{
L_{\text{max}} := \text{max} \br{L_+, L_-} {= L_{-}}.
\label{Lmax}
}
The {important} point is the existence of a critical length $L_c$ in the region of $M^2 > 1/L_{\text{max}}^2$, whose definition is
\al{
L_c = \frac{1}{|M|} \text{arctanh} \paren{ \frac{|M|\paren{L_+ + L_-}}{1 + M^2 L_+ L_-} }
\qquad \text{for } M^2 > \frac{1}{L_{\text{max}}^2},
}
and when we introduce a $U(1)$ gauge boson, the gauge symmetry is broken (unbroken) for $L < L_c$ ($L \geq L_c$). On the other hand, the gauge symmetry is always broken in the region of $M^2 \leq \frac{1}{L_{\text{max}}^2}$.

Now we discuss properties of the solution.
{With} the condition of
\al{
k \sim 1 \quad \text{and} \quad -\sqrt{\frac{\lambda}{2}} \frac{\mu}{k} y_0 \gg 1,
}
the form of the solution gets to be exponential as follows:
\al{
\vev{\Phi(y)} \underbrace{\sim}_{k\sim1} \nu \cosh\paren{\sqrt{\frac{\lambda}{2}} {\mu} (y-y_0)}
\underbrace{\sim}_{\sqrt{\frac{\lambda}{2}} {\mu} (y-y_0) \gg 1}
\frac{\nu}{2} e^{-\sqrt{\frac{\lambda}{2}} {\mu} y_0} \cdot
e^{\sqrt{\frac{\lambda}{2}} {\mu} y},
\label{approximatedHiggsVEV}
}
which is just the form of the {{warped}} VEV.
{Detailed analysis,} including numerical calculation{, is provided in Section~\ref{sec:Detailed_analysis}}.
For {a} more concrete understanding, we {conduct further analysis based on} the discussion in Ref.~\cite{Fujimoto:2011kf}.
We introduce the eigenfunctions $f_{(n)}(y)$ of the eigenvalue equation
\al{
-\pal_y^2 f_{(n)}(y) = E_{(n)} f_{(n)}(y), \qquad n=0,1,2,\cdots,
}
with the {BCs}
\al{
f_{(n)}(0) + L_+ \pal_y f_{(n)}(0) = f_{(n)}(L) - L_- \pal_y f_{(n)}(L) = 0.
}
In terms of the orthonormal eigenfunctions $f_{(n)}(y)$, whose orthonormality is ensured by the hermiticity of the operator $(-\pal_y^2)$, the field $\Phi$ can be expanded as
\al{
\Phi(y) = \sum_{n=0}^{\infty} \phi_{(n)} f_{(n)}(y),
}
with the corresponding coefficients $\phi_{(n)}$.
Inserting {this} into $\E[\Phi]$ in Eq.~(\ref{Higgseffectivepotential}) leads to
\al{
\E[\Phi] = \sum_{n=0}^{\infty} m_{(n)}^2 \ab{\phi_{(n)}}^2 + \paren{\text{quartic terms in }\phi_{(n)}},
}
where
\al{
m_{(n)}^2 := M^2 + {E_{(n)}}, \qquad n=0,1,2,\cdots.
}
Note that the quartic terms are non-negative for any configurations of $\phi_{(n)}$ because they come from the term of $\int_0^L dy \frac{\lambda}{2} \ab{\Phi}^4\,(\geq 0)$.
{It follows} that the configuration of the VEV is given by $\vev{\Phi} = 0$ $(\text{or }\vev{\phi_{(n)}} = 0\text{ for any }n)$ if $m_{(n)}^2 \geq 0$ for any $n$.
For realizing {symmetry breaking}, the condition $m_{(0)}^2 < 0$ is mandatory and the $\ab{\phi_{(0)}}^2$ term's contribution is probably dominant around the minima.\footnote{
Continuing discussions in this way, we can classify the phase structure of the {scalar} without knowing the details of the VEV $\vev{\Phi}$.
}
Consequently, we could approximate the form of $\vev{\Phi(y)}$ as
\al{
\vev{\Phi(y)} \sim \phi_{(0)} f_{(0)}(y).
\label{approximatedPhifield}
}
Here we take the two parameters of the BC as
\al{
\frac{1}{L_{\pm}} = \mp \paren{M + \epsilon},
}
where $\epsilon$ is {a} microscopic (but not infinitesimal) positive value. 
It is not difficult to obtain the form of $f_{(0)}(y)$ and the corresponding eigenvalue $E_{(0)}$ with the assumptions
\al{
f_{(0)}(y) \sim e^{(M+\epsilon)y}, \qquad  {E_{(0)}} = -(M+\epsilon)^2,
}
{from which} we derive the form of the {{warped}} VEV again.
By calculating the value of $m_{(0)}$ as
\al{
{m_{(0)}^2} = - 2 \epsilon M - \epsilon^2,
}
we can infer {with certainly that} symmetry breaking is realized with our premise of $M>0$.\footnote{
In other {words}, the condition $M^2 < 1/L_{\text{max}}^2$ is fulfilled.
}

%%%%%%%%%%%%%%%%%%%%%%%%%%%%%%%%%%%%%%%%%%%%%
\subsection{A model for realizing our mechanism without violating gauge universality in {multiple} point interaction {systems}}
%%%%%%%%%%%%%%%%%%%%%%%%%%%%%%%%%%%%%%%%%%%%%

Now we know that the form of the {{warped}} VEV can be achieved by the {scalar} with the generalized BC in Eq.~(\ref{RobinBC}).
{Thinking naively}, in the UED-type model with one Higgs $SU(2)_W$ {\it doublet} $H$ with the generalized BCs at the two end {points} of the total system $(y=0,y={L})$, and {``continuous"} conditions at the others, we can explain the quark mass hierarchy and the structure of the CKM matrix simultaneously via {geometry}.
But two nontrivial issues exist in this setup.

In the SM, the profile of the VEV $\vev{H}$ can be rotated by use of {$SU(2)_W$} global symmetry as
\al{
\vev{H} \rightarrow \begin{pmatrix} 0 \\ {v}/\sqrt{2L} \end{pmatrix},
\label{rotatedHiggsVEV}
}
with $v = 246\,\text{GeV}${,} because the profile $\vev{H}$ is constant.
The first issue is {whether} we can rotate the profile and obtain the above form or not.
In our case{, the VEV generally} becomes $y$-dependent and the 4D effective Higgs potential in Eq.~(\ref{Higgseffectivepotential}) is considered to hold a very complicated structure.
Thus it is nontrivial {whether} the VEV $\vev{H}$ can take the form in Eq.~(\ref{rotatedHiggsVEV}).

The second issue is {critical}.
When the VEV is $y$-position dependent, {the} zero mode profile of a 4D gauge boson is determined by a Lam\'{e}-type equation and is not constant any longer.
Then the overlap integrals for quark-antiquark-gauge boson interactions in the SM become generation-dependent and{,} as a consequence, the gauge universality in the SM is jeopardized.
Of course{,} this result is unacceptable and {therefore} we need to alter our strategy.

A remedy for this {problem requires} the coexistence of {a Higgs doublet $H$ and a singlet scalar $\Phi$}, where the BC of the former is chosen as the ordinary Neumann-type at the end points{,}
\al{
\pal_y H|_{y=0} = \pal_y H|_{y={L}} = 0,
\label{doubletBCs}
}
and of the latter is selected {as} the generalized BC in Eq.~(\ref{RobinBC}).
The 5D actions for $H$ and $\Phi$ are given as
\al{
S_{H} &= \int d^4 x \int_0^{{L}} dy \bigg\{ {H^{\dagger} (D_M D^M + {M'}^2 ) H}
		- \frac{\lambda'}{2} \paren{H^\dagger H}^2 \bigg\}, 
\label{H_action} \\
S_{\Phi} &= \int d^4 x \int_0^{{L}} dy \bigg\{ \Phi^{\dagger} \paren{\pal_M \pal^M - {M}^2} \Phi
		- \frac{\lambda}{2} \paren{\Phi^\dagger \Phi}^2 \bigg\}.
\label{Phi_action}
}
Here we introduce the 5D gauge bosons {$G_M$, $W_M$, $B_M$}, which are 5D $SU(3)_C$, $SU(2)_W$, $U(1)_Y$ {bosons}, respectively.
$D_M$ means the covariant derivative for the corresponding gauge bosons.
The 5D fermions in Eq.~(\ref{effectiveaction}) are also gauged for reproducing the SM interactions.
$D_M$ and {the} action for the 5D gauge bosons take the same form {as} those in the {mUED} and we do not {include them here} since the detailed information is not important in the following discussion.
The BCs for {$G_M$, $W_M$, $B_M$} are {chosen} as
\al{
\pal_y G_\mu|_{y=0} = \pal_y G_\mu|_{y={L}} &= 0, \\
G_y|_{y=0} = G_y|_{y={L}} &= 0,
}
where we only illustrate the gluon case.
{Fig.~\ref{BosonBCprofile_pdf} shows} a schematic diagram for explaining the BCs for bosons. The green, orange{,} and purple circular spots represent the ordinary Neumann, Dirichlet{,} and the generalized BCs in Eq.~(\ref{RobinBC}), respectively.
{Note that except for} the both end points of the total system, we {use} the continuous condition discussed in Eqs.~(\ref{gaugebosoncondition3_twointerval}) and (\ref{gaugebosoncondition4_twointerval}).
{Each 4D vector part has} a zero mode, whose mode function is a constant.
By {using the condition} ${{M'}^2 > 0}$, {$SU(2)_W \times U(1)_Y$} gauge symmetry is spontaneously broken through the usual Higgs mechanism and the SM situation is duplicated
at the sectors of {$G_M$, $W_M$, $B_M$, $H$}.
Here we should assign the $U(1)_Y$ charge of the Higgs singlet $\Phi$ as zero to {ensure} that $\Phi$ {does} not couple to any gauge bosons.
Accordingly, the problem {with} gauge universality never occurs in the refined setup.
\begin{figure}
\centering
\includegraphics[width=0.7\columnwidth]{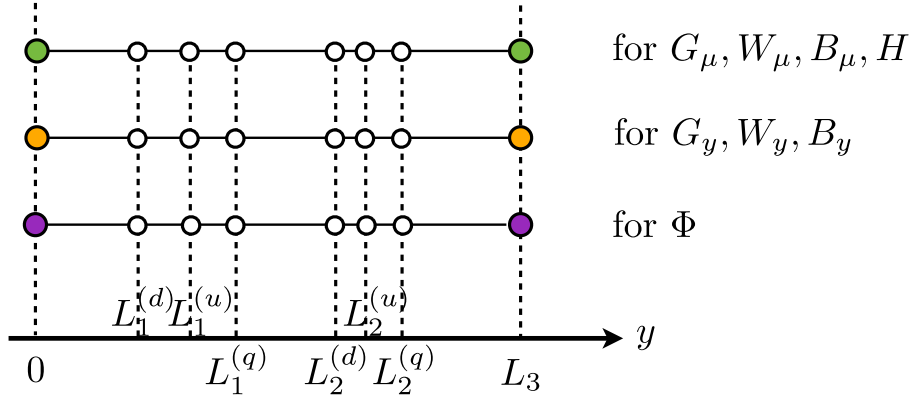}
\caption{
A schematic diagram for explaining the BCs for bosons.
The green, orange {and} purple circular spots represent the ordinary Neumann, Dirichlet{,} {and} the generalized {BCs} in Eq.~(\ref{RobinBC}), respectively. {Note that except for} the both end points of the total system, we {use} the continuous condition discussed in Eqs.~(\ref{gaugebosoncondition3_twointerval}) and (\ref{gaugebosoncondition4_twointerval}).
}
\label{BosonBCprofile_pdf}
\end{figure}

What we should {consider} next is the structure of 5D Yukawa interactions.
When we adopt the forms in the SM (or the {mUED}), the large mass hierarchy cannot be created since the profile of $\vev{H}$ is constant.
To simplify the situation, we introduce {the discrete symmetry}
\al{
H \rightarrow - H, \qquad \Phi \rightarrow - \Phi
\label{Higgsdiscretesymmetry}
}
to prohibit the terms of {$\overline{Q} (i \sigma_2 H^{\ast}) \U$, $\overline{Q} H \D$, $\Phi \overline{Q} Q$, $\Phi \overline{\U} \U$, $\Phi \overline{\D} \D$} with the Pauli matrix $\sigma_2$.
The desirable 4D Yukawa structure is generated by introducing the {terms}
\al{
S_{\Y} = \int d^4 x \int_{0}^{{L}} dy \bigg\{ \Phi {\Big[
		- \Y^{(u)} \overline{Q} (i \sigma_2 H^{\ast}) \U - \Y^{(d)} \overline{Q} H \D \Big]
		+ \text{h.c.}}
		\bigg\},
\label{higherdimensionalYukawa}
}
where those operators are higher-dimensional compared to the previous five operators and allowed under the discrete symmetry in Eq.~(\ref{Higgsdiscretesymmetry}).
We {note that} the coefficients {$\Y^{(u)}$, $\Y^{(d)}$} have mass dimension $-2$.
After the {electroweak symmetry breaking} (EWSB) occurs {with nonvanishing $\vev{H}$ and $\vev{\Phi}$}, {the} 4D effective Lagrangian which we assume in Eq.~(\ref{effectiveaction}) is realized without any serious {conflict} with the nature of the SM.

{{Now} we discuss some related issues. Whether 5D gauge invariance is conserved or not is one of {the important criteria} for judging validity of BCs. When we break gauge symmetry by BCs, the issue of possible unitary violation due to longitudinal components of massive gauge bosons should be considered~\cite{Csaki:2003dt,Sakai:2006qi,Nishiwaki:2010te}.
{However, we} mention that in the cases of the Dirichlet BC for the fermions at {the} mid and end points and the generalized Higgs BC at {the} end {points}, 5D gauge invariance is intact.}

In the above analysis, we neglect the contribution to the total scalar potential of the doublet-singlet mixing term with {coefficient} $C${:}
{
\al{
S_{\text{mixing}} = \int d^4 x \int_{0}^{{L}} dy \bigg\{ -C \Phi^\dagger \Phi H^\dagger H \bigg\},
\label{doubletsingletmixingterm}
}
where the discrete symmetry in Eq.~(\ref{Higgsdiscretesymmetry}) cannot {proclude} {this form}.
After considering this part, the profiles of $\Phi$ and $H$ are deformed and the problem {with} gauge universality is revived.
Therefore we should choose a sufficiently small coefficient $C$ to avoid this obstacle.
{Detailed discussion of this topic is provided} in Appendix~\ref{Appendix1} {and \ref{Appendix2}}.}

%%%%%%%%%%%%%%%%%%%%%%%%%%%%%%%%%%%%%%%%%%%%%
\subsection{Detailed numerical calculations for justifying the model with Higgs doublet and singlet
\label{sec:Detailed_analysis}}
%%%%%%%%%%%%%%%%%%%%%%%%%%%%%%%%%%%%%%%%%%%%%

Based on the previous discussions, we {re-examine the issue of the validity of} our model including the Higgs doublet and {scalar} singlet with numerical calculations.
At first, we reconsider the approximation in Eq.~(\ref{approximatedHiggsVEV}).
As we have discussed before, the $\text{cn}$ function is almost equivalent to the $\text{cosh}$ function in the limit of $k\simeq1$, and then we obtain the form
\al{
\vev{\Phi(y)} \simeq \nu \cosh\sqbr{\sqrt{\frac{\lambda}{2}} \mu (y-y_0)} =
\frac{\nu}{2} \bigg\{ e^{\sqrt{\frac{\lambda}{2}} \mu (y-y_0)} + e^{- \sqrt{\frac{\lambda}{2}} \mu (y-y_0)} \bigg\} \qquad (\text{with } k\simeq1).
\label{approximatedHiggsVEV_1}
} 
{If} the condition $\sqrt{\frac{\lambda}{2}} \mu (y-y_0) \gtrsim 1$ is fulfilled, the second {term} on the {right-hand side} of Eq.~(\ref{approximatedHiggsVEV_1}) can be neglected {with} about $10\%$ error $(e^{-2} \simeq 0.135)$.
{Therefore,} we get the outcome of
\al{
\vev{\Phi(y)} \simeq \frac{\nu}{2} e^{\sqrt{\frac{\lambda}{2}} \mu (y-y_0)}
		\qquad \paren{\text{with } k \simeq 1 \text{ and } \sqrt{\frac{\lambda}{2}} \mu (y-y_0) \gtrsim 1}.
\label{approximatedHiggsVEV_2}
}
Here we consider the situation of $k \simeq 1$ more concretely.
As we show in Eq.~(\ref{parameter_k}), {the parameter $k$} is composed {from} some input parameters for the solution{,} and $k \simeq 1$ is {equivalent} to the condition
\al{
X = \frac{4\lambda \ab{Q}}{M^4} \simeq 0.
}
It is obvious that for matching this condition, smaller (greater) {values} of $\lambda$ and/or $\ab{Q}$ $(M)$ {are} preferred.
But the extremal choices of $\lambda=0$, $\ab{Q}=0$, $M=\infty$ turn into disorder and unnaturalness.
We rewrite the approximated form in Eq.~(\ref{approximatedHiggsVEV_2}) with input parameters by considering the shapes of {$\mu$, $\nu$} in Eqs.~(\ref{form_mu}) and (\ref{form_nu}), which are 
{approximately,} under the situation $k \simeq 1${,}
\al{
\mu \simeq \sqrt{\frac{2}{\lambda}} M, \quad \nu \simeq \frac{\sqrt{2\ab{Q}}}{M},
}
with the zeroth (first) order approximation in $X$ for $\mu$ $(\nu)$.
When we evaluate $\nu$ up to $X$'s zeroth order, the value of $\nu$ goes to zero and this is meaningless. 
{Using these} results, we can rewrite the equation in (\ref{approximatedHiggsVEV_2}) as
\al{
\vev{\Phi(y)} \simeq \sqrt{\frac{\ab{Q}}{2}}\frac{1}{M} e^{-My_0} \cdot e^{My}
		\qquad \paren{\text{with } k\simeq1 \text{ and } M (y-y_0) \gtrsim 1}.
\label{approximatedHiggsVEV_3}
} 
The correspondence {of Eq.~(\ref{approximatedHiggsVEV_3})} to the assumed {warped} VEV in Eq.~(\ref{assumedHiggsVEVprofile}) is as follows:
\al{
\mathcal{A} = \sqrt{\frac{\ab{Q}}{2}}\frac{1}{M} {e^{M(L-y_0)}}, \quad
		\alpha = M.
\label{Higgssingletcorrespondence}
}
Then we notice that the shape of the approximated profile in Eq.~(\ref{approximatedHiggsVEV_3})  is mainly determined by {$M$ and $y_0$}, and that $\lambda$ does not appear {in} the approximation.

In what follows, we discuss the {validity} of the above approximation and the deviation from it when we consider the exact form in Eq.~(\ref{exactHiggsVEVform}).
At first, we define the {dimensionless} {parameters} with tilde \,$\tilde{}$\,  in the basis of the massive parameter of the total length of the system $L\,(=L_3)$, {e.g.},
\al{
M = \tilde{M} L^{-1}, \quad L_i^{(q)} = \tilde{L}_i^{(q)} L,
\label{scaledM}
}
where {some} of the {dimensionless} parameters {are} already calculated in Eq.~(\ref{a_setofsolution2}).
The significant point is {that} the bulk mass of the Higgs singlet is already almost fixed because of Eqs.~(\ref{a_setofsolution2}), (\ref{approximatedHiggsVEV_3}){,} and (\ref{Higgssingletcorrespondence}) as
\al{
{M \simeq 8.67 L^{-1}, }
\label{scaledMvalue}
}
and {so} we should search for a region of the parameters related to the singlet under {this} constraint.

In our configuration, the modulus parameter of Jacobi's elliptic function $\text{cn}$ is determined as a function of {$X = \frac{4 \lambda \ab{Q}}{M^4}$,} as in Eq.~(\ref{parameter_k}){,} and the relation between them is shown in Fig.~\ref{ellipticcn1and2_pdf}, where we understand that we have to take very small $X$ {to obtain} $k=1$.
We also refer to the complete elliptic integral of the first kind $K[k]$, which is {a} function of the elliptic modulus $k$ and whose value is equal to the quarter period of Jacobi's elliptic function $\text{cn}(y,k)$ in Fig.~\ref{ellipticcn1and2_pdf}.
This plot suggests that if we take the infinite period, {which} corresponds to $1/\text{cosh}(y,k)$, we tune the value of $k$ very close to one.
\begin{figure}
\centering
\includegraphics[width=0.9\columnwidth]{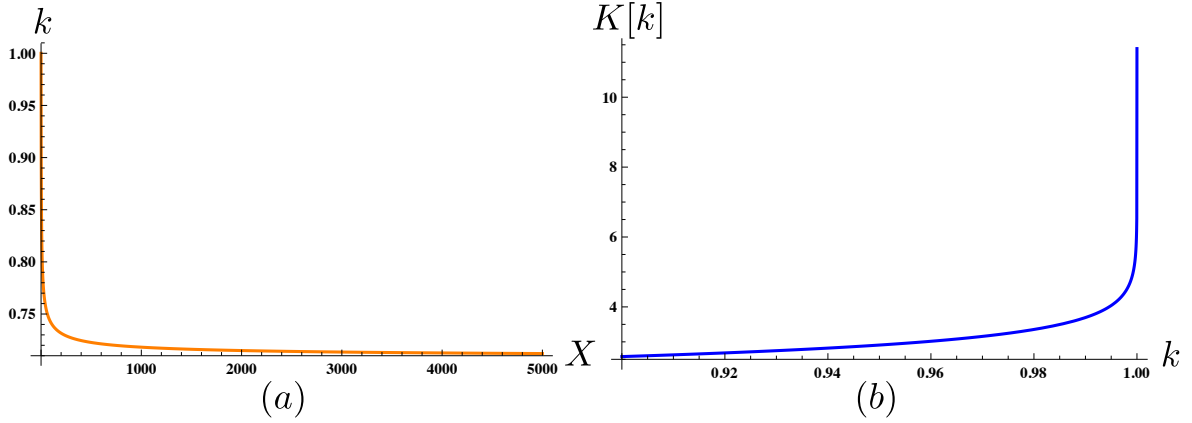}
\caption{
{(a)} {The} value of $k$ in the function of $X$ in the range of $0 \leq X \leq 5000$.
{(b)} {The} value of the complete elliptic integral of the first kind $K[k]$ in the range of $0.9 \leq k \leq 1$.
}
\label{ellipticcn1and2_pdf}
\end{figure}

{The positions of the divergent points} of $1/\text{cn}(x,k)$, which {correspond to zero points} of  $\text{cn}(x,k)$, give us another important suggestion.
The $1/\text{cn}(x,k)$ function {gets} divergent with a period of $2K[k]$ and the range of $[0,L]$ should not contain any such point.
In the exact form of the VEV $\vev{\Phi(y)}$ in Eq.~(\ref{exactHiggsVEVform}), {the position $y_d$ with divergence} is evaluated as
\al{
\tilde{y}_d = \tilde{y}_0 + \frac{1}{\tilde{M} \paren{1+X}^{1/4}} {\textstyle K\sqbr{ \sqrt{ \frac{1}{2} \paren{1 + \frac{1}{\sqrt{1+X}}} } } } \qquad \paren{\text{mod } \frac{2}{\tilde{M} \paren{1+X}^{1/4}} {\textstyle K\sqbr{ \sqrt{ \frac{1}{2} \paren{1 + \frac{1}{\sqrt{1+X}}} } } } }
\label{divergentpoint}
}
$\text{with } X=\tilde{X}= \frac{4 \tilde{\lambda} |{\tilde{Q}}|}{\tilde{M}^4}$.
{The second term of the {right-hand} side in Eq.~(\ref{divergentpoint}) means the quarter period of $\vev{\Phi(y)}$ in the coordinate $\tilde{y}$.}
Considering the profile in Fig.~\ref{VEVplots_pdf}, in the scaled coordinate $\tilde{y}$ the position of $\tilde{y}_d$ is preferred at one plus a few positive values.
When we consider the property of the scaled quarter period in Fig.~\ref{ellipticcn3and4_pdf},
as we have discussed before,
we need to make the value of $X$ {approach} zero (but not {exactly} zero) for {an} {$\mathcal{O}(1)$ scaled period}.
In addition, we have to take account of the condition on $y_0$ in Eq.~(\ref{approximatedHiggsVEV_3}),
which is interpreted in the scaled coordinate $\tilde{y}$ as
\al{
{
\tilde{M} (\tilde{y} - \tilde{y}_0) \gtrsim 1 \rightarrow 8.67 (\tilde{y} - \tilde{y}_0) \gtrsim 1
\qquad \text{in the range of } 0 \leq \tilde{y} \leq 1, }
}
with the value of dimensionless $\tilde{M}$ in Eq.~(\ref{scaledMvalue}).
Here we observe that the case of positive $\tilde{y}_0$ is problematic (at least) around $\tilde{y} = 0$.
Based on all the knowledge which we have obtained, we can find a set of parameters{:}
\al{
{\tilde{M} = 8.67}, \quad \tilde{y}_0 = - 0.1, \quad \tilde{\lambda} = 0.001, \quad |\tilde{Q}| = 0.001.
\label{asetofHiggsparameters}
}
The validity of this choice is checked in Fig.~\ref{ellipticcn3and4_pdf} to calculate the ratio, which is defined as the exact VEV form over its approximated form in Eq.~(\ref{approximatedHiggsVEV_3}), and the difference is estimated as about {$15\%$} at most.
%Some tuning in the Higgs singlet's parameters is hard to avoid since the elliptic function $\text{cn}(y,k)$ get to be close to the exponential function only in limited and extremal case of $k$.
%
%
\begin{figure}
\centering
\includegraphics[width=0.99\columnwidth]{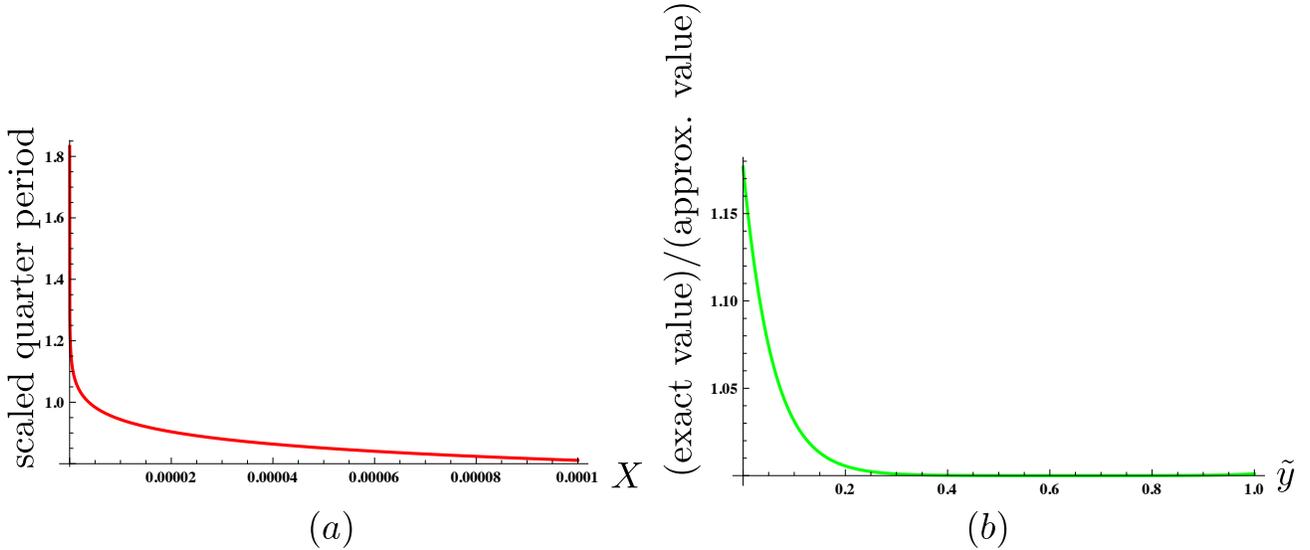}
\caption{
{(a)} {The} value of the scaled quarter period of the exact VEV in the function of $X$ in the range of $0 \leq X \leq 0.0001$.
{(b)} {The} value of the ratio, which is defined as the exact VEV form over its approximated form in Eq.~(\ref{approximatedHiggsVEV_3}).
}
\label{ellipticcn3and4_pdf}
\end{figure}

In the following, we check whether the {EWSB} occurs or not in our configuration.
The prescription is written in Section~\ref{Position-dependent Higgs VEV in generalized boundary condition} and the two input parameters $L_{\pm}$ can be inversely calculated from the profile of the exact VEV through Eq.~(\ref{inverseLs}) as
\al{
{
\frac{1}{\tilde{L}_+} = - 6.07, \quad \frac{1}{\tilde{L}_-} = 8.69 }
\label{Lplusminusvalues}
}
and the value of $\tilde{L}_{\text{max}}$ in Eq.~(\ref{Lmax}) is {simultaneously} fixed {at} {$1/\tilde{L}_{\text{max}} = 8.69$} as the scaled values based on $L$.
Agreeing with the previous naive discussion at the end of Section~\ref{Position-dependent Higgs VEV in generalized boundary condition}, the condition
\al{
M^2 < \frac{1}{L_{\text{max}}^2}
}
is fulfilled{, and therefore} the EWSB is realized in our configuration for real.
Here we comment on two things.
One is that the parameter $\lambda$ always appears in the $\text{cn}$ function as the combination $|Q|\lambda${,} and $\lambda$ in itself only affects the overall normalization.
This means that we can take {greater values of} $\lambda$ with {smaller} $|Q|$.
The other is that the smallness of $y_0$ and $|Q|$ is not {an} unnatural thing, because they are resultant values derived from the two input parameters $L_{\pm}$, whose {dimensionless} values are within $\mathcal{O}(10)$.

{From} the above discussion, we are able to calculate {the} Yukawa mass matrix elements in our model with {the} ``elliptic VEV" $\vev{\Phi(y)}$.
We take care of the following two facts:
\begin{itemize}
\item We use the exact form of the VEV in Eq.~(\ref{exactHiggsVEVform}) instead of the assumed {{warped}} form in Eq.~(\ref{assumedHiggsVEVprofile}) with the parameters in Eqs.~(\ref{asetofHiggsparameters}) and/or (\ref{Lplusminusvalues}).
\item Because the Yukawa structure is introduced as the higher-dimensional {operators} in Eq.~(\ref{higherdimensionalYukawa}), {their} replacement is required
\al{
Y^{(u)} \rightarrow \Y^{(u)} \frac{v}{\sqrt{2}}, \quad
Y^{(d)} \rightarrow \Y^{(d)} \frac{v}{\sqrt{2}}.
\label{Yukawareplacement}
}
\end{itemize}
The diagonalized Yukawa mass matrices take the forms
\al{
\mathcal{M}^{(u)}|_{\text{diagonal}} &= {\text{diag}\, (2.47\,\text{MeV},\ 1.18\,\text{GeV},\ 174\,\text{GeV})}, \\
\mathcal{M}^{(d)}|_{\text{diagonal}} &= {\text{diag}\, (3.94\,\text{MeV},\ 110\,\text{MeV},\ 4.19\,\text{GeV})},
}
and the CKM matrix is given as
\al{
{|V_{\text{CKM}}| =}
\begin{bmatrix}
		0.977 &  0.214 &  0.00448 \\
		 0.213 &  0.976 & 0.0475 \\
		0.0145 &  0.0454 & 0.999
\end{bmatrix}, 
}
where we adopt the values in Eq.~(\ref{a_setofsolution2}) for the 9 lengths {($L_1^{(q)}$, $L_2^{(q)}$, $L_3^{(q)}$, $L_1^{(u)}$, $L_2^{(u)}$, $L_3^{(u)}$, $L_1^{(d)}$, $L_2^{(d)}$, $L_3^{(d)}$)} and the 3 fermion bulk masses {(${M_Q}$, $M_\U$, $M_\D$)}.
We can find only a small difference between the results and the previous results based on the assumed {warped} VEV.

In contrast to the foregoing analysis, the Higgs dynamics is well described and the dimensionless coefficient part of $\vev{\Phi(y)}$, which is related to $\tilde{\A}$ in the assumed {{warped}} VEV, is calculable and the values of {$\tilde{\Y}^{(u)}$, $\tilde{\Y}^{(d)}$} are as follows:
\al{
\tilde{\Y}^{(u)} = 0.0442, \quad
\tilde{\Y}^{(d)} = 0.00369.
}
Here we {note} that when we consider the situation where all the objects are localized, the effective length of each object is considered to be smaller than the total length $L$. 
If we take the effective length for overlap integrals showing down quarks as about {$L/3$}, which is the length of the integral for $m_{11}^{(d)}$, 
the unnaturalness in $\tilde{\Y}^{(d)}$ is somewhat ameliorated as follows:
\al{
\tilde{\Y}^{(d)} \sim 0.00369 \cdot (3)^2 = 0.0332.
}
{The insertion of the relatively-large massive value $v=246\,\text{GeV}$, compared to the quark masses {(except the top quark), in} the 5D Yukawa structure in Eq~(\ref{Yukawareplacement}) is a cause of the smallness of $\tilde{\Y}^{(u)}$ and $\tilde{\Y}^{(d)}$.}
We do not {provide} more detailed discussion since we cannot avoid some ambiguities.

%%%%%%%%%%%%%%%%%%%%%%%%%%%%%%%%%%%%%%%%%%%%%
%%%%%%%%%%%%%%%%%%%%%%%%
\section{Summary and discussions
\label{Summary_section}}
%%%%%%%%%%%%%%%%%%%%%%%%
%%%%%%%%%%%%%%%%%%%%%%%%%%%%%%%%%%%%%%%%%%%%%

We have presented a review {of} some properties of 5D fermion and vector fields in {a} multi-interval system, where each interval is connected to the others through {point interactions}.
By choosing {suitable} BCs at the positions of {the point interactions}{,} the profiles of {the fermion} zero modes get to be three-fold degenerated, localized, and mixed, which means all of the Yukawa structure in the SM can be realized.
Combined with the $y$-position-dependent Higgs VEV with exponential {warped} form, the quark mass hierarchy and the structure of the CKM matrix are explained simultaneously almost only via {the geometry of the extra dimension}.
One way to generate the {warped} VEV without gauge universality violation is to introduce both the Higgs doublet with the ordinary Neumann BCs and the scalar singlet with the generalized BCs, which are coupled in higher-dimensional Yukawa terms. The ordinary Yukawa terms are prohibited by adding a discrete symmetry.
The exact form of the scalar singlet VEV is {represented by} Jacobi's elliptic function and we have found that it becomes close to the exponential function in a region of parameters with almost $\mathcal{O}(10)$ input parameters.
To avoid violation in gauge universality, we should assume that the magnitude of the coefficient of the doublet-singlet mixing term in Eq.~(\ref{doubletsingletmixingterm}) is sufficiently small.

Here we {briefly} estimate the effect from KK mixing.
In our system, translational invariance along $y$ is highly violated because of the existence of the point interactions{,} and moreover KK-parity cannot be defined {because of the} lack of reflection symmetry.
Consequently, the zero modes and KK modes of the fermions are mixed at the tree level and this {affects the values} of the mass eigenstates and the elements of the CKM matrix.
Here we would like to consider the form
\al{
-
\begin{bmatrix}
\overline{t}^{(0)},\ \overline{T}^{(1)},\ \overline{t}^{(1)}
\end{bmatrix}_L
\begin{bmatrix}
m_t & 0 & m_t \\
m_t & M_{\text{KK}} & m_t \\
0 & m_t & {-M_{\text{KK}}}
\end{bmatrix}
\begin{bmatrix}
t^{(0)} \\
T^{(1)} \\
t^{(1)}
\end{bmatrix}_R
+ \text{h.c.},
}
where {$m_t$, $M_{\text{KK}}$, $t^{(0)}$, $T^{(1)}$, and $t^{(1)}$} are the top quark mass, KK scale, top zero mode, first top KK state of {the} $SU(2)_W$ doublet, {and that of the} $SU(2)_W$ singlet, respectively.
In the following, we only consider mixing in the top quark sector {since} mixings in the other five flavors are negligible because of {the} smallness of their Yukawa couplings.
In reality{,} some deviation factors from the assumed ordinary UED-like form probably emerge in front of each matrix component{, originating from differences} in overlap integrals and mixings {between a zero mode and/or between zero modes and} KK states of more than {the} first level are also derived.
But we ignore these issues for simplicity.

The deviation ratio in the observed top mass, which is defined as the mass eigenvalues over the reference value {$173.3 \,(\pm 2.8) \,\text{GeV}$, whose value is from a recent work~\cite{Alekhin:2012py}, is calculated and the {result is} represented in Fig.~\ref{levelmixing_pdf}.
The yellow band shows the allowed region after considering the error {of the observed top quark mass}.}
\begin{figure}
\centering
\includegraphics[width=0.65\columnwidth]{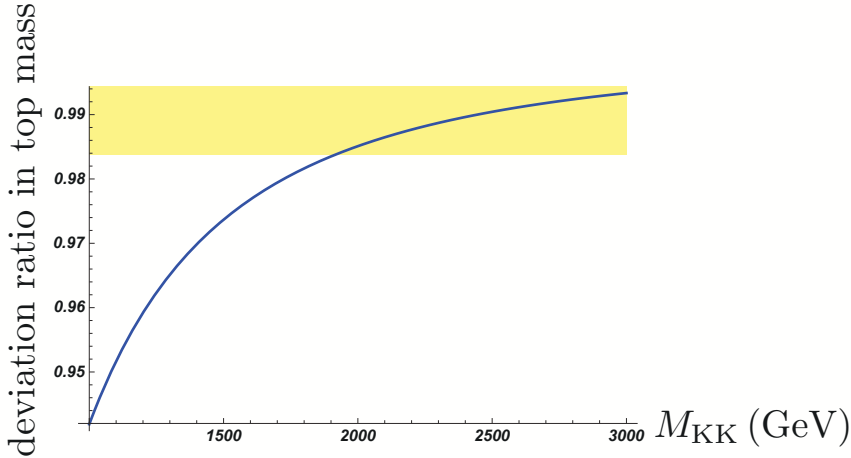}
\caption{
The deviation ratio in the observed top mass, which is defined as the mass eigenvalues over the reference value {$173.3 \,(\pm 2.8) \,\text{GeV}$, whose value is from a recent work~\cite{Alekhin:2012py}.
The yellow band shows the allowed region after considering the error {of the observed top quark mass}.}
}
\label{levelmixing_pdf}
\end{figure}
{We} conclude that we can ignore the level-mixing effect when the KK scale, which is defined as $\pi/L$ in the ordinary UED context, is greater than {$2\,\text{TeV}$}{, dependent on} accepting our naive assumptions.
More {detailed} analysis is out of the scope of this paper and we will discuss this issue in a future study.
It is noted that in a ``decoupling" case with a huge $M_{\text{KK}}$, we can always neglect the {level-mixing} effect even if this possibility is not so interesting {from} {a} collider physics point of view.

The work done in this paper is considered as a first step for constructing a {phenomenological} model which explains
the number of fermion {generations}, fermion mass hierarchies, and the structure of fermion mixing matrices simultaneously{, based mainly on the} geometry of an extra dimension.
But lots of issues {remain to be studied}.

The first issue is whether our mechanism works well in the lepton sector.
As {is} widely known, two mixing angles of neutrinos are large, and this suggests that the components of {the} neutrino mass matrix are probably the same order of magnitude{,}
in contrast to the quark one.
In addition, the expected neutrino masses are up to {(sub-)}$\text{eV}$ order, {whereas} we find at least six orders of magnitude smaller than the value of the lightest charged lepton (electron).
These differences are nontrivial and we will search for a configuration where the properties of quarks and leptons are derived simultaneously via geometry.
The second related issue is on the phase of CP violation in the CKM matrix.
The existence of this phase with nonzero value has been established by B physics experiments,
but within our {present} mechanism such a phase never occurs since all the zero mode functions are real.
{One possible direction for overcoming} this problem is to introduce {a complex phase through} twisted boundary conditions.

{Away from phenomenological issues}, the Higgs with the generalized boundary conditions has a rich theoretical structure{, and} many topics wait {there to be} unveiled.
The phase structure of the scalar singlet is explored in Ref.~\cite{Fujimoto:2011kf}{,} but for non-singlet scalars {only} limited studies have been done.
In non-Abelian gauge theory the boundary conditions can mix the gauge indices and {the number of possibilities is increased. Accordingly,} its phase structure gets to be much richer.
Another interesting issue is the structure of quantum fluctuation around the position-dependent elliptic VEV of the singlet scalar.
Even in the case of the zero mode, the properties are highly nontrivial but this issue is mandatory when we discuss the signature of the scalar singlet at colliders.
Moduli stabilization via Casimir energy is also an important {topic} for ensuring the stability of the system with the nontrivial VEV structure.\footnote{
We can find some related works in Refs.~\cite{de_Albuquerque:2003uf,Bajnok:2005dx,Pawellek:2008st}.
}

The physics in the system with {point interactions} and/or $y$-position-dependent {scalar} VEV {is only starting to be discovered,} we can find a lot of fascinating themes {from} both phenomenological and theoretical points of view.

%%%%%%%%%%%%%%%%%%%%%%%%%%%%%%%%%%%%%%%%%%%%%%%%%%%%%%%
%%%%%%%%%%%%%%%%%%%%%%%%%%%%%%%%%%%%%%%%%%%%%%%%%%%%%%%
%%%%%%%%%%%%%%%%%%%%%%%%%%%%%%%%%%%%%%%%%%%%%%%%%%%%%%%
%\vskip 10pt
%\noindent
%{\large \bf Acknowledgments}
%\vskip 7pt
%\noindent
\section*{Acknowledgment}

The authors appreciate S.Ohya for giving us many fruitful comments and reading our manuscript carefully.
The authors would like to thank T.Kugo, N.Maru, K.-y.Oda, J.Sato, Y.Shimizu, R.Takahashi, M.Yamanaka, and T.Yamashita for valuable discussions.
K.N. is partially supported by funding available from the Department of 
Atomic Energy, Government of India for the Regional Centre for Accelerator-based
Particle Physics (RECAPP), Harish-Chandra Research Institute.
This work is supported in part by a Grant-in-Aid for Scientific Research (No. 22540281 and No. 20540274 (M.S.)) from the Japanese Ministry of Education, Science, Sports and Culture.

%%%%%%%%%%%%%%%%%%%%%%%%%%%%%%%%%%%%%%%%%%%%%%%%%%%%%%%%%%%%%%%
%%%%%%%%%%%%%%%%%%%%%%%%%%%%%%%%%%%%%%%%%%%%%%%%%%%%%%%%%%%%%%%
%%%%%%%%%%%%%%%%%%%%%%%%%%%%%%%%%%%%%%%%%%%%%%%%%%%%%%%%%%%%%%%
%%%%%%%%%%%%%%%%%%%%%%%%%%%%%%%%%%%%%%%%%%%%%%%%%%%%%%%%%%%%%%%

\appendix
\section*{Appendix}

%%%%%%%%%%%%%%%%%%%%%%%%%
\section{Estimating the orders of magnitude of doublet-singlet {scalar} mixing {effects} \label{Appendix1}}
%%%%%%%%%%%%%%%%%%%%%%%%%
In this Appendix, we consider the minimization problem of the scalar potential of the doublet $H$ and singlet $\Phi$ scalars with the doublet-singlet mixing term in Eq.~(\ref{doubletsingletmixingterm}). In this analysis, we assume that the VEV of the singlet $\Phi$ takes the effective form $\vev{\phi(y)}$ {from} Eq.~(\ref{assumedHiggsVEVprofile}) and we concentrate on the part of
\al{
\int d^4 x \int_0^{L} dy \bigg\{ {H^{\dagger} (\pal_y)^2 H} + {M'}^2 H^\dagger H
		- \frac{\lambda'}{2} \paren{H^\dagger H}^2
		- C \Phi^\dagger \Phi H^\dagger H \bigg\}.
}
After the replacement
\al{
H \rightarrow \begin{pmatrix} 0 \\ \frac{\vev{h(y)}}{\sqrt{2}}  \end{pmatrix}, \quad
\Phi \rightarrow \vev{\phi(y)},
}
we can identify the functional form $\E'[\vev{h}]$ which we should minimize as follows:
\al{
\E'[\vev{h}] = 
		\int_0^{L} dy \bigg\{ (\pal_y \vev{h})^2 - {M'}^2 \vev{h}^2
		+ \frac{\lambda'}{4} \vev{h}^4
		+ C \vev{\phi}^2 \vev{h}^2 \bigg\}.
}

Due to the Neumann BCs in Eq.~(\ref{doubletBCs}), the form of the (position-dependent) VEV $\vev{h(y)}$ is fixed as
\al{
\vev{h(y)} = \sqrt{\frac{2}{\lambda'}} M' + \beta_0 + \sum_{n=1}^{\infty} \beta_n \cos\paren{\frac{\pi n}{L} y},
\label{twoscalar_functional}
}
with the coefficients {$\beta_0$, $\beta_n$}.
Here the first term of the {right-hand} side of Eq.~(\ref{twoscalar_functional}) corresponds to the solution with $C=0$, whose value is equal to ${v/\sqrt{L}}$ with $v=246\, \text{GeV}$, and the remaining two terms show the deformation from it in the case of $C \not= 0$.
Under the assumption that {the value of $C$ is small}, the potential is minimized with the forms of the coefficients
\al{
\beta_0 &=  - \frac{\sqrt{\frac{2}{\lambda'}} C}{2 {M'} L} \frac{{\A^2}}{2\alpha} \paren{1-e^{-2\alpha L}}, \label{beta_zero} \\
\beta_n &=  - \frac{{2} \sqrt{\frac{2}{\lambda'}} M' C}{L \sqbr{2{M'}^2 + \paren{\frac{\pi n}{L}}^2}} \frac{{\A^2}}{2} \frac{4 \alpha}{4 \alpha^2 + \paren{\frac{\pi n}{L}}^2} \paren{(-1)^n -e^{-2\alpha L}} \label{beta_n},
}
within the second order of $C$.

After we consider the suitable order estimation
\al{
\lambda' \sim L, \quad M' \sim v, \quad \A \sqrt{L} \sim v, \quad L^{-1} \sim M_{\text{KK}}, \quad \alpha L \sim \mathcal{O}(1),
\label{orderestimation}
}
where $M_{\text{KK}}$ is a typical mass scale of the KK states,
we can conclude that the orders of magnitude of the deviation from $C=0$ {are}
\al{
\ab{\frac{\beta_0}{\sqrt{\frac{2}{\lambda'}} M'}} &\sim \frac{\tilde{C}}{\tilde{\alpha}}, \\
\ab{\frac{\beta_n}{\sqrt{\frac{2}{\lambda'}} M'}} &\sim \frac{\tilde{\alpha}}{n^2 \paren{n^2 + \tilde{\alpha}^2}} \paren{\frac{v}{M_{\text{KK}}}}^2 \tilde{C},
}
with the {dimensionless} values {$\alpha = \tilde{\alpha}L^{-1}$, $C = \tilde{C} L$}.
The {results} tell us that the value of $\beta_n$ is suppressed by the KK index $n$ and $M_{\text{KK}}$, but{,} on the contrary, that of $\beta_0$ is not suppressed {because} $\tilde{\alpha} = \mathcal{O}(1)$ in Eqs.~(\ref{a_setofsolution2}) and (\ref{orderestimation}){,}
although $\beta_{0}$ does not {affect} our
conclusions since it merely shifts the constant expectation value
of $H$. On the other hand, nonzero values of $\beta_{n}$ could
cause a problem {for} gauge universality, so that $\tilde{C}$
({$M_{\text{KK}}$}) should be sufficiently small (large) in our model.

%%%%%%%%%%%%%%%%%%%%%%%%%%%%%%%%%%%%%%%%%%%%%%%%%%%%%
%%%%%%%%%%%%%%%%%%%%%%%%%%%%%%%%%%%%%%%%%%%%%%%%%%%%%
%\section{}
%%%%%%%%%%%%%%%%%%%%%%%%%%%%%%%%%%%%%%%%%%%%%%%%%%%%%
%%%%%%%%%%%%%%%%%%%%%%%%%%%%%%%%%%%%%%%%%%%%%%%%%%%%%

%%%%%%%%%%%%%%%%%%%%%%%%%
\section{Evaluating {the} upper bound of the coefficient of {the} doublet-singlet mixing term via gauge universality violation in Z boson decay branching ratios\label{Appendix2}}
%%%%%%%%%%%%%%%%%%%%%%%%%

Following the previous Appendix, we evaluate an upper bound {for} the coefficient of doublet-singlet mixing term $\tilde{C}$ via gauge universality violation in Z boson decay branching ratios.
As {is} widely known, the value of {the} decay width of {a} Z boson into quarks and leptons is generation-independent {(taking} the massless limit on the fermions in the final state) and its possible deviation is considered to be a good order parameter for gauge universality violation in our model.
{In this analysis, we only focus on the case of $\tilde{C} \geq 0$ since in the case of $\tilde{C} < 0$, we also consider {the} stability condition of the total scalar potential.
The full potential analysis is considered to be an interesting {topic} and we will leave it for a future work.}

First we focus on the zero mode physical Higgs part of the Higgs doublet $H${:}
\al{
H \rightarrow \begin{pmatrix} 0 \\ \frac{1}{\sqrt{2}} \left( \vev{h} + h^{(0)} \right) \end{pmatrix},
}
where $\vev{h}$ ($h^{(0)}$) is the VEV (quantum fluctuation of its zero mode), respectively.
Due to {the} variational principle, {the} equation for determining {the profile $f_{h^{(0)}}$ of $h^{(0)}$} is derived as follows:
\al{
\left( -\partial_{y}^{2} - M'^{2} + \frac{3}{2} \lambda' \vev{h(y)}^{2} + C \langle \phi(y) \rangle^{2} \right) f_{h^{(0)}}(y) = \mu_{h^{(0)}}^2 f_{h^{(0)}}(y),
}
where $\mu_{h^{(0)}}$ is the physical mass of $h^{(0)}$ and the forms of $\vev{h}$ and $\langle \phi \rangle$ {have} been already discussed in Appendix~\ref{Appendix1}.
We solve the equation with the BC of $\partial_y f_{h^{(0)}}|_{y=0} = \partial_y f_{h^{(0)}}|_{y=L} = 0$ up to {a} perturbation of first order of $\tilde{C}$.

After dividing $\vev{h}$, $f_{h^{(0)}}${,} and $\mu_{h^{(0)}}$ {into} their unperturbed and perturbed parts
\al{
\vev{h} = {v_5^{(0)}} + \varphi^{(1)},\quad
f_{h^{(0)}} = f_{h^{(0)}}^{(0)} + f_{h^{(0)}}^{(1)}, \quad \mu_{h^{(0)}}^2 = {(m^{(0)}_{h^{(0)}})^2 + (m^{(1)}_{h^{(0)}})^2}, \label{perturbed_values}
}
where ${v_5^{(0)}} = \sqrt{2M'^2/\lambda'} = v/\sqrt{L}$ is the 5D unperturbed $h$'s VEV{,} the upper indices $``(0)"$ and $``(1)"$ show the order of perturbation {with respect to $\tilde{C}$.}
%and {$(m^{(0)}_{h^{(0)}})^2$ ($(m^{(1)}_{h^{(0)}})^2$)} is the unperturbed (perturbed) squared mass, respectively.
{The} concrete forms are determined via the BCs and orthonormality as follows:
\al{
f_{h^{(0)}}^{(0)}(y) &= \frac{1}{\sqrt{L}},\quad {(m^{(0)}_{h^{(0)}})^2} = 2M'^2, \\
f_{h^{(0)}}^{(1)}(y) &= \frac{1}{\sqrt{L}} \Bigg[ A_{h^{(0)}} + B_{h^{(0)}} y 
   + 3 \lambda' {v_5^{(0)}} \left( \beta_0 \frac{y^2}{2} - \sum_{n=1}^{\infty} \beta_n \left( \frac{L}{n \pi}
   \right)^2 \cos\left( \frac{n\pi}{L} y \right) \right) \notag \\
   &\phantom{= \frac{1}{\sqrt{L}} \Bigg[ } -\frac{1}{2} {(m^{(1)}_{h^{(0)}})^2}\, y^2
   + C \mathcal{A}^2 \frac{1}{(2\alpha)^2} e^{2\alpha(y-L)} \Bigg], \label{perturbed_Higgsprofile} \\
{(m^{(1)}_{h^{(0)}})^2} &= 3\lambda' {v_5^{(0)}} \beta_0 + \frac{C\mathcal{A}^2}{2\alpha L}
   \left( 1-e^{-2\alpha L} \right), \label{Delta_Higgs}
}
where {$\beta_0$ and $\beta_n$ are already shown in Eqs.~(\ref{beta_zero}) and (\ref{beta_n}), respectively,} and the coefficients for Eq.~(\ref{perturbed_Higgsprofile}) {are given by}
\al{
A_{h^{(0)}} &= \frac{C \mathcal{A}^2 L}{12\alpha}  \left( 1+2e^{-2\alpha L} \right)   
   - \frac{C \mathcal{A}^2}{(2\alpha)^3 L} \left( 1-e^{-2\alpha L} \right), \\
B_{h^{(0)}} &= - \frac{C\mathcal{A}^2}{2\alpha} e^{-2\alpha L}.
}

Next, we also {estimate} {the} perturbed Z boson profile $f_{Z^{(0)}}$ and physical mass $\mu_{Z^{(0)}}$.
The corresponding EOM takes the form\footnote{
As discussed in \cite{Haba:2009wa}, gauge-fixing terms should be introduced to eliminate some mixing terms.}
\al{
\left( -\partial_y^2 + {(m^{(0)}_{Z^{(0)}})^2} + \sqrt{g_5^2 + g_5'^2} \, {m^{(0)}_{Z^{(0)}}}
   \varphi^{(1)}(y) \right) f_{Z^{(0)}}(y) = \mu_{Z^{(0)}}^2 f_{Z^{(0)}}(y).
}
Here{,} $g_5$ ($g_5'$) is the 5D $SU(2)_W$ ($U(1)_Y$) gauge coupling and ${m_{Z^{(0)}}^{(0)}} = \frac{1}{2} {v_5^{(0)}} \sqrt{g_5^2 + g_5'^2}$ is the unperturbed Z boson mass.
As in the case of the physical Higgs, we can obtain the perturbed results with the same notation for order of perturbation as
\al{
f_{Z^{(0)}}^{(0)}(y) &= \frac{1}{\sqrt{L}}, \\
f_{Z^{(0)}}^{(1)}(y) &= - \frac{1}{\sqrt{L}} \sum_{n=1}^{\infty} \beta_n \left( \frac{L}{n\pi} \right)^2
   \sqrt{g_5^2 + g_5'^2} \, {m^{(0)}_{Z^{(0)}}} \cos\left( \frac{n\pi}{L}y \right), \label{perturbed_Zbosonprofile} \\
{(m^{(1)}_{Z^{(0)}})^2} &= \sqrt{g_5^2 + g_5'^2}\, {m_{Z^{(0)}}^{(0)}} \beta_0.
   \label{Delta_Zboson}
}
As expected, the perturbed Z boson profile in Eq.~(\ref{perturbed_Zbosonprofile}) is $y$-dependent and probably becomes a source of gauge universality violation.

From Eqs.~(\ref{perturbed_values}), (\ref{Delta_Higgs}) and (\ref{Delta_Zboson}), we can obtain the following relations
\al{
\tilde{\lambda'}\ {(:= \lambda'/L)} &= \frac{1}{4} \left( g^2 + g'^2 \right) 
   \left(\frac{\mu_{h^{(0)}}}{\mu_{Z^{(0)}}}\right)^2, \label{relation1}\\
{(m^{(0)}_{Z^{(0)}})^2} &= \mu_{Z^{(0)}}^2 + \frac{1}{4} \left( g^2 + g'^2 \right) 
   \frac{\tilde{\mathcal{A}}^2}{\tilde{\lambda'} \tilde{\alpha} \pi^2} \tilde{C} M_{\text{KK}}^2,
   \label{relation2}
}
where $g = g_5/\sqrt{L}$ and $g' = g'_5/\sqrt{L}$ are the corresponding 4D gauge couplings, and $M_{\text{KK}}$ is defined as $\pi/L$ in this Appendix.
{Note that we ignored the small factor of $e^{-2 \alpha L}$ in Eqs.~(\ref{relation1}) and (\ref{relation2}), where $\alpha L = 8.67$ was assigned in Eqs.~(\ref{Higgssingletcorrespondence}) and (\ref{asetofHiggsparameters}).}
Before {proceeding to} numerical {calculations}, we {summarize} some important issues:
\begin{itemize}
\item
The overall Z boson perturbed profile is
\al{
f_{Z^{(0)}}(y) = \frac{1}{\sqrt{L}} \left[ 1 + \sum_{n=1}^{\infty} (-1)^n 8 \tilde{C}
   {(m^{(0)}_{Z^{(0)}})^2} \frac{\tilde{\mathcal{A}}^2}{n^2 \pi^2 + 2 \tilde{M'}^2}
   \frac{\tilde{\alpha}}{n^2 \pi^2 + 4 \tilde{\alpha}^2}
   \frac{1}{n^2 M_{\text{KK}}^2} {\cos\left( \frac{n\pi}{L}y \right)} \right],
   \label{fullZbosonprofile}
}
where {the effects from higher modes $(n \geq 2)$ are found to be numerically suppressed} and we can ignore them.
\item
The relation between Higgs quartic coupling and physical mass in Eq.~(\ref{relation1}) is the
same {as in} the SM even after the perturbation.
\item
The relation between unperturbed and perturbed physical masses in Eq.~(\ref{relation2}) indicates that when we take the value of $M_{\text{KK}} \gg \mu_{Z^{(0)}}\,(\simeq 90\,\text{GeV})$, ${(m^{(0)}_{Z^{(0)}})^2}$ is simply expressed with good precision as
\al{
{(m^{(0)}_{Z^{(0)}})^2} \sim M_{\text{KK}}^2.
}
This means that {the} upper bound on $\tilde{C}$ probably gets to be (almost) constant in the range of $M_{\text{KK}}$ above a few TeV{,} and the effect of {the} $n=1$ mode in Eq.~(\ref{fullZbosonprofile}) is never decoupled even in the limit of $M_{\text{KK}} \rightarrow \infty$.
%\item
%If the unperturbed Z-boson squared mass $m_{Z^{(0)}}^2$ gets to be negative, the Z-boson becomes tachyonic, in other words, the unperturbed theory is destabilized and the whole discussion turns into meaningless.
%Considering the potential hierarchy of $M_{\text{KK}} > \mu_{Z^{(0)}}$, $\tilde{C}$ with very small absolute value is only acceptable in the case of $\tilde{C} < 0$.
%In this paper, we do not pursue this possibility and concentrate on cases with $\tilde{C} \geq 0$.
\end{itemize}

\begin{table}[t]
   \centering
  \begin{tabular}{|c|ccc||ccc|} \hline
    type & $a$ & $b$ & $f_{fL}$ & $c$ & $d$ & $f_{fR}$ \\ \hline \hline
    up & $0$ & $L_1^{(q)}$ & $f_{q_{1L}^{(0)}}$ & $0$ & $L_1^{(u)}$ & $f_{u_{1R}^{(0)}}$ \\
    down & $0$ & $L_1^{(q)}$ & $f_{q_{1L}^{(0)}}$ & $0$ & $L_1^{(d)}$ & $f_{d_{1R}^{(0)}}$ \\ 
    strange & $L_1^{(q)}$ & $L_2^{(q)}$ & $f_{q_{2L}^{(0)}}$ & $L_1^{(d)}$ & $L_2^{(d)}$ & $f_{d_{2R}^{(0)}}$ \\
    charm & $L_1^{(q)}$ & $L_2^{(q)}$ & $f_{q_{2L}^{(0)}}$ & $L_1^{(u)}$ & $L_2^{(u)}$ & $f_{u_{2R}^{(0)}}$ \\
    bottom & $L_2^{(q)}$ & $L$ & $f_{q_{3L}^{(0)}}$ & $L_2^{(d)}$ & $L$ & $f_{d_{3R}^{(0)}}$ \\
    \hline
  \end{tabular}
\caption{Summary table for left- and {right-handed} couplings of quarks to the Z boson.}
\label{integration_parameters}
\end{table}

\begin{figure}[t]
\centering
\includegraphics[width=0.99\columnwidth]{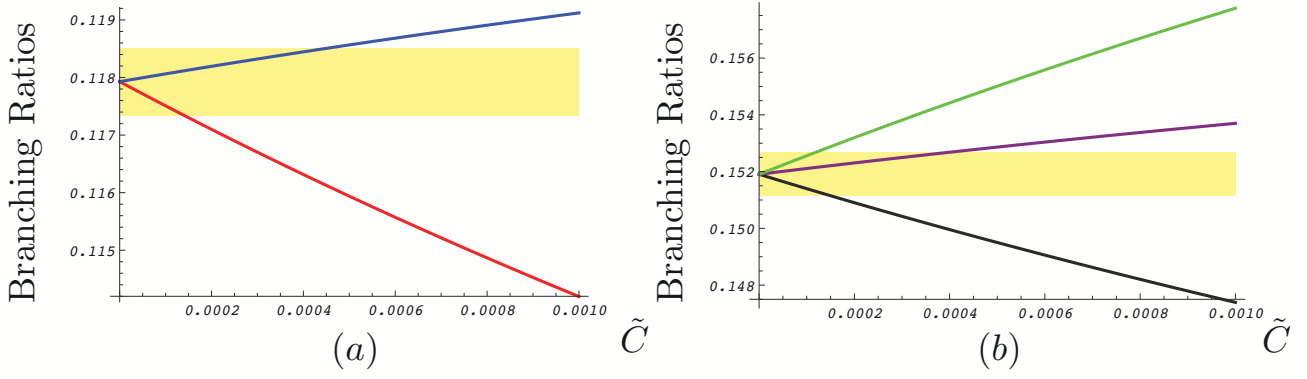}
\caption{The Z boson's branching ratios with $M_{\text{KK}} = 100\,\text{GeV}$.
The red and blue curves in $(a)$ represent {branching} into a pair of up and charm quarks, while in $(b)$ the black, purple{,} and green curves show {branching} into a pair of down, strange{,} and bottom quarks, respectively.
The horizontal yellow band shows the allowed region, which is defined as {where} the deviation from the SM is within $0.5\%$.
}
\label{GUV100GeV_pdf}
\end{figure}
\begin{figure}[t]
\centering
\includegraphics[width=0.99\columnwidth]{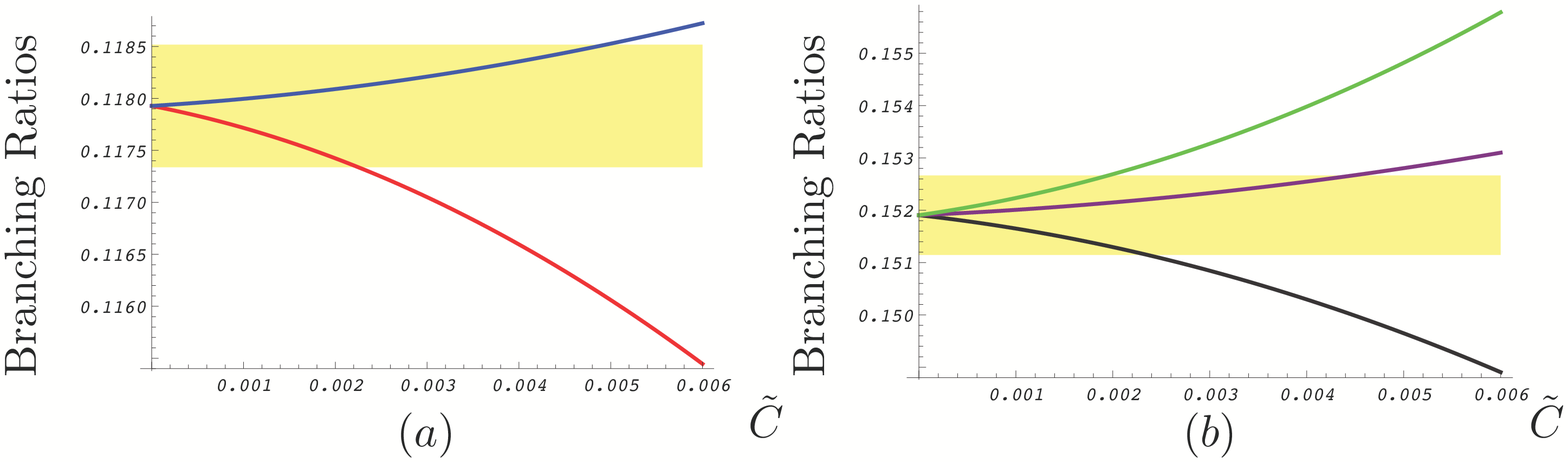}
\caption{The Z boson's branching ratios with $M_{\text{KK}} = 500\,\text{GeV}$.
The color code is {as} in Fig.~\ref{GUV100GeV_pdf}.
}
\label{GUV500GeV_pdf}
\end{figure}
\begin{figure}[t]
\centering
\includegraphics[width=0.99\columnwidth]{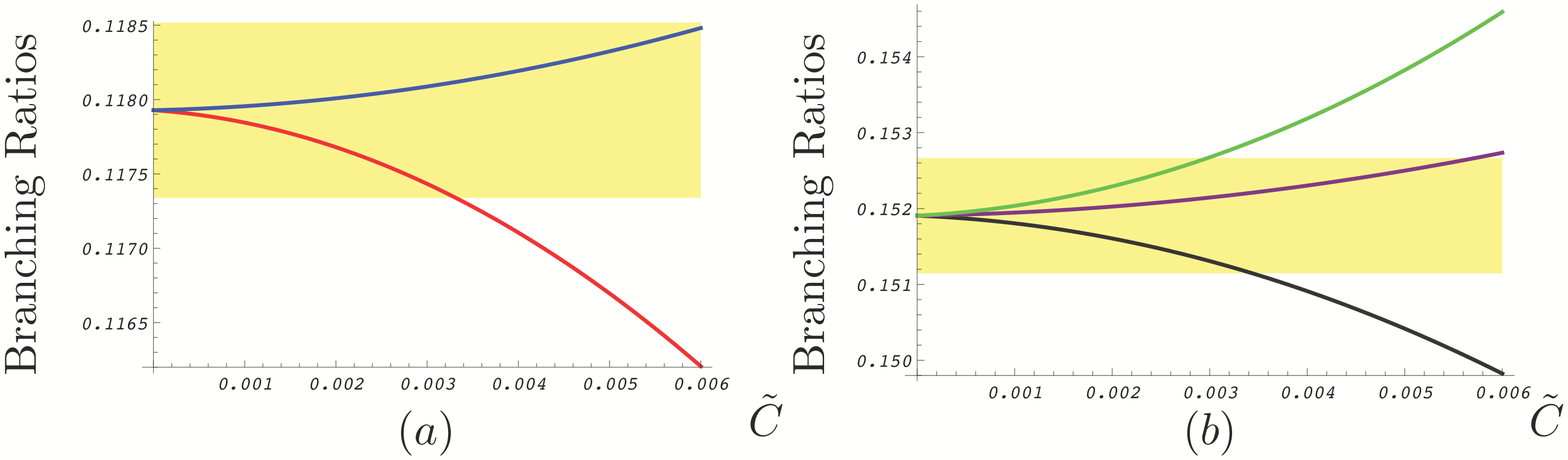}
\caption{The Z boson's branching ratios with $M_{\text{KK}} = 1\,\text{TeV}$.
The color code is {as} in Fig.~\ref{GUV100GeV_pdf}.
}
\label{GUV1TeV_pdf}
\end{figure}
\begin{figure}[t]
\centering
\includegraphics[width=0.99\columnwidth]{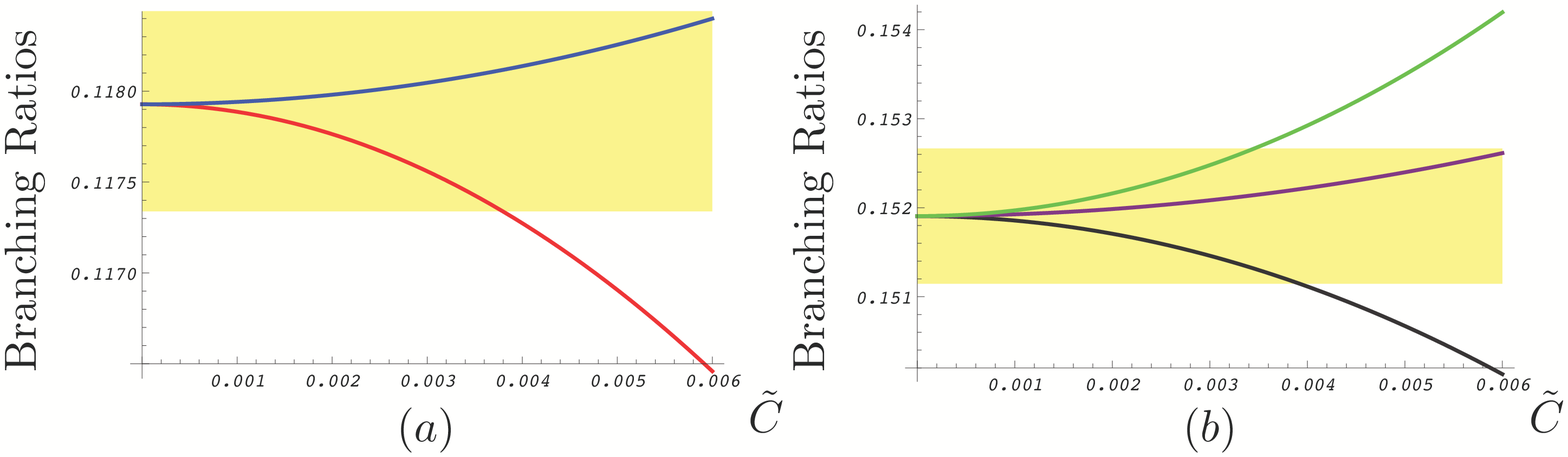}
\caption{The Z boson's branching ratios with $M_{\text{KK}} = 10\,\text{TeV}$.
The color code is {as} in Fig.~\ref{GUV100GeV_pdf}.
}
\label{GUV10TeV_pdf}
\end{figure}

In our model, partial widths of the Z boson are evaluated by the formula
\al{
\Gamma(Z^{(0)} \rightarrow \overline{f^{(0)}} f^{(0)}) = 4 N_C \left[ (G_{fL})^2 + (G_{fR})^2 \right] \Gamma^{0}_{Z},
}
where $N_C$ is {the} QCD color facor ($3$ for quark, $1$ for lepton) and $\Gamma^0_Z = G_F (\mu_{Z^{(0)}})^3/{12\sqrt{2}\pi}$ with the Fermi constant $G_F$.
Here we {assume} that the fermions in the final {state} are massless.
For a quark with its profiles {$\{ f_{fL}\,\text{(left)}, f_{fR}\,\text{(right)} \}$}, the left- {(right-)handed} coupling $G_{fL}$ ($G_{fR}$) is represented in general as follows:
\al{
G_{fL} &= g_{fL} \int_{a}^{b} dy f_{fL}^2 \left( \sqrt{L} f_{Z^{(0)}} \right), \\
G_{fR} &= g_{fR} \int_{c}^{d} dy f_{fR}^2 \left( \sqrt{L} f_{Z^{(0)}} \right),
}
where $g_{fL}$ ($g_{fR}$) corresponds to the values in the SM, whose form is{:}
\al{
g_{fL}^2 = (I^3_{W,f})^2 - 2 \sin^2{\theta_W} I^3_{W,f} Q_f + \sin^4{\theta_W} Q_f^2,\quad g_{fR}^2 &= \sin^4{\theta_W} Q_f^2,
}
with {the} third component of weak isospin ($I^3_{W,f}$), the Weinberg angle ($\theta_W$), and electromagnetic charge in a unit of the elementary charge ($Q_f$).
The input values for each integration are summarized in Table~\ref{integration_parameters}.
We note that in the limit $\tilde{C} \rightarrow 0$, the form of $f_{Z^{(0)}}$ gets back to the unperturbed one ($=1/\sqrt{L}$) and then $G_{fL}$ and $G_{fR}$ become generation-independent due to {the} orthonormality of the fermion profiles.
On the other hand{,} for leptons, whose profiles are {out} of the scope of this paper, we assume {that} there is no point interaction in the bulk space and consequently their left- and right-handed couplings are {entirely the same as in} the SM.

In {our} numerical {calculations}, we adopt the values which we used in Section~\ref{concretemodel_section} and we set the physical Higgs mass as $125\,\text{GeV}$.
We consider the four possibilities of $M_{\text{KK}}${,} $100\,\text{GeV}$, $500\,\text{GeV}$, $1\,\text{TeV}${,} and $10\,\text{TeV}$ in Figs~\ref{GUV100GeV_pdf}, \ref{GUV500GeV_pdf}, \ref{GUV1TeV_pdf}{, and \ref{GUV10TeV_pdf} respectively}.
In all the plots, the red, blue, black, purple{,} and green curves represent the decay branching ratios of the Z boson into a pair of up, charm, down, strange{,} and bottom quarks, respectively.
The horizontal yellow {bands show} the allowed {regions, which are} defined as {where} the deviation from the SM is within $0.5\%$, {which is the} typical accuracy of the latest experimental data~\cite{Beringer:1900zz}. 
Here we can recognize two things.
One is {that} in the every {case}, deviations in the bottom quark put the most stringent bound on $\tilde{C}$.
The other is {that} when we take $M_{\text{KK}}$ as above $1\,\text{TeV}$, the upper bound on $\tilde{C}$ is almost the same as we anticipated before.
To conclude, when we choose a TeV-scale value in $M_{\text{KK}}$, where this choice is preferable in terms of phenomenological consistency, the value of $\tilde{C}$ should be located in the range of
\al{
{\tilde{C} \lesssim 0.003.}
}

%%%%%%%%%% References %%%%%%%%%%%

\begin{comment}

\end{comment}

%\bibliographystyle{TitleAndArxiv}
%\bibliographystyle{JHEP}
%\bibliography{maintextref}

\bibliographystyle{utphys}
\bibliography{sakamoto_paper}

%\bibliographystyle{TitleAndArxiv}
%\bibliography{letter}

\end{document}